\input psfig.tex
\input apj.macro
\overfullrule=0pt
%
%

\def \gta {\mathrel{\vcenter
          {\hbox{$>$}\nointerlineskip\hbox{$\sim$}}}} 

 \def\etal{{et al.} \thinspace}

 \def\:{\mskip\medmuskip}                         
%

%
%
 \def\ni{$^{56}{\rm Ni}\ $}      
%
 \def\kelvin{\thinspace\rm{\sp{o}{\kern-.08333em }K}\ }
%
%
%
\title{Type Ia Supernovae: Influence of the Initial Composition }
\title{on the Nucleosynthesis, Light Curves,  Spectra}
\title{and Consequences for the Determination of  $\Omega _M$ \& $\Lambda$}
\bigskip
\bigskip
\centerline{P.~H\"oflich$^{1,2,3}$, J.C. Wheeler $^{1,3}$, F.K. Thielemann
$^{2,3}$}
\bigskip
\leftline{1. Department of Astronomy, University of Texas, Austin, TX 07871,
USA}
\leftline{2. Department of Physics and Astronomy, University of Basel, CH-4056
Basel, Switzerland}
\leftline{3. Institute for Theoretical Physics, University of California, Santa
Barbara,
CA 93186-4030, USA}
\bigskip
\bigskip

\heading{Abstract}

\noindent
The influence of the initial composition of the exploding white dwarf on
the nucleosynthesis,  light curves
and spectra of  Type Ia  supernovae
 has been studied in order to evaluate the size of evolutionary effects
on cosmological time scales, how the effects can be recognized and how one may
be able to correct for
them.

 The calculations are based on a set of delayed detonation models which give a
good account of
the optical and infrared light curves  and
of the spectral evolution.
 The explosions and light curves are calculated using a one-dimensional
Lagrangian radiation-hydro
code  including  a nuclear network.
Spectra are computed
for various epochs  using the
structure resulting from the light curve code. Our NLTE code
 solves the relativistic radiation transport equations in the
 comoving frame  consistently  with the
statistical equations and ionization due to $\gamma $ radiation
for the most important elements (C, O, Ne, Na, Mg, Si, S, Ca, Fe, Co, Ni).
About  1,000,000 additional lines  are included
 assuming LTE-level populations and an equivalent-two-level approach for the
source functions.

 Changing the initial metallicity Z from Population I to II
alters the isotopic composition of the outer layers of the ejecta that have
undergone explosive O
burning.
 Especially important is the
 increase of  the $^{54}$Fe production with metallicity.
The influence on the resulting rest frame visual and blue light
curves is found to be small.
 Detailed analysis of spectral evolution should permit a determination of the
progenitor metallicity.

Mixing $^{56} Ni$ into the outer layers during the explosion
can produce  effects similar to an increased initial metallicity.
 Mixing can be distinguished from metallicity effects
   by means of the
  strong cobalt and nickel lines, by  a change of the calcium lines in the
optical and IR spectra
and, in principle, by $\gamma $-ray observations.

 As the C/O ratio of  the WD is decreased, the explosion energy and
 the \ni production are reduced and the Si-rich layers are more confined in
velocity space. A reduction of C/O by about 60 \% gives  slower
rise times by about 3 days, an increased luminosity at maximum light,
 a somewhat faster post-maximum decline and
a larger ratio between maximum light and \ni tail.
 A reduction of the C/O ratio has a similar effect on the colors, light curve
shapes
and element distribution as a  reduction in the deflagration to detonation
transition
density but, for the same light curve shape, the absolute brightness
 is larger for smaller C/O.
 An independent determination of  the initial C/O ratio and the transition
density
 is possible for local SN if detailed analyses of both the spectra and
light curves are performed simultaneously.

Because        the spectra are shifted into different color bands at different
redshifts,   the  effect of metallicity Z on a given observed color
 is  a strong function of redshift.
 A change    of Z by a factor of 3  or the C/O ratio by 33 \%
alters the  peak magnitudes in the optical wavelength range by
up to  $\approx 0.3^m$ for $z \geq 0.2$. These variations are comparable to the
effect of
changes of $\Omega_M$ and $\Lambda$ at redshifts of 0.5 to 1.
The  systematic effects due to  changes in composition are expected to remain
small up
to about $z \approx 0.5$ for R-V and up to $z\approx 0.7$ for R-I.

With proper account
of evolutionary corrections,
supernovae will provide a valuable tool  to determine the cosmological
parameters and they will provide  new insight into the
chemical evolution of the universe.

\bigskip \noindent
{\it Subject headings:} Supernovae and supernovae remnants: general --
hydrodynamics --
nucleosynthesis -- radiation transfer  -- light curves -- $H_o$,  $\Omega _M$,
$\Lambda$
%
%
%
\section{Introduction}

Type Ia Supernovae (SNe~Ia)
may reach the same brightness as the entire parent galaxy. In
 principle, this allows for the measurements of extragalactic distances
 and
cosmological parameters such as the Hubble constant $H_o$, $\Omega_M$,
$\Omega_\Lambda$,
 and the related deceleration
parameter $q_o$.
It is now widely accepted
that Type Ia supernovae are not a strictly homogeneous class of objects with
similar brightness
(e.g. Barbon, Ciatti \& Rosino, 1978; Pskovskii 1970, 1977;  Branch 1981;
Phillips et al. 1987,
Filippenko \etal 1992a; Phillips \etal 1992; Filippenko \etal 1992ab;
Leibundgut \etal 1993;
 Hamuy \etal  1996).
For nearby supernovae ($z \leq 0.1$),
 different schemes have been developed and tested
to cope with the problems of deducing the intrinsic brightness
 based on theoretical models or observed correlations between spectra or light
curves and the
absolute brightness using primary distance indicators
(e.g. Norgaard-Nielsen et al. 1989;
 Branch \& Tammann 1992, , Sandage \& Tammann 1993,
 M\"uller \& H\"oflich 1994,
Hamuy et al. 1996; Riess, Kirshner \& Press 1995; Nugent et al. 1995;
 H\"oflich \& Khokhlov
1996). These methods have been tested
locally and provide consistent results.
 New telescopes and observational
 programs put SNe~Ia at large red shifts well within reach and justify
optimism for the discovery of a large number of distant SNe~Ia.
The Berkeley group has discovered more than 50 SNe~Ia
up to a red shift of 0.9 (Pennypacker et al. 1991, Perlmutter et al. 1997;
Pennypacker, private communication). The
 CfA/CTIO/ESO/MSSSO collaboration  has found a similar number of
supernovae (Leibundgut et al. 1995,  Schmidt et al. 1996).
 Systematic errors due to evolutionary effects
represent a concern for  the use of SNe~Ia to determine the shape of the
universe
(H\"oflich et al. 1997). This is especially
true for statistical methods  which are calibrated only on local SNe~Ia.
 This  paper represents a first attempt to characterize some of these
evolutionary effects and
points the way to eliminate associated systematic errors.

 It is  widely accepted that SNe~Ia are thermonuclear
explosions of carbon-oxygen white dwarfs (Hoyle \& Fowler 1960; for discussions
of various theoretical aspects see Woosley \&  Weaver 1986, 1995,
Wheeler \& Harkness 1990, Canal 1994, Nomoto et al. 1995, Nomoto 1995,
Wheeler et al. 1995, and H\"oflich \&  Khokhlov 1996). Three main scenarios
can be distinguished:

 A primary scenario consists of massive carbon-oxygen white
dwarfs (WDs) with a mass close to the Chandrasekhar mass which accrete
through Roche-lobe overflow from an evolved companion star (Nomoto \&
Sugimoto 1977; Nomoto 1982).  In these accretion models, the explosion is
triggered by compressional heating.
{}From the theoretical standpoint, the key questions are
how the flame ignites and propagates through the white
dwarf. Several models of SNe~Ia within this general scenario have been
proposed in the past, including detonations (Arnett 1969; Hansen \& Wheeler
1969),
deflagrations (Ivanova, Imshennik \& Chechetkin 1974; Nomoto, Sugimoto \&
Neo 1976) and  delayed detonations,
which assume that the flame starts as a deflagration and turns into a
detonation later on (Khokhlov 1991,  Yamaoka \etal 1992,
Woosley \& Weaver 1995). The latter scenario, the so-called ``delayed
detonation"  and
its variation ``pulsating delayed detonation", seems to be the most promising
one,
because,  from the general properties and the individual
light curves and spectra, it  can  account for the majority
of SNe~Ia events (e.g. H\"oflich \& Khokhlov 1996,
 H\"oflich et al. 1997, Nomoto et al. 1997, Nugent et al. 1997,
and references therein).
 In addition, with the discovery of the supersoft X-ray sources,
potential progenitors have been found (van den Heuvel et al. 1992; Rappaport et
al. 1994ab;
DiStefano et al. 1997).
We note that the classical
``deflagration'' model W7 has a similar structure to the ``delayed detonation"
models
which have been successfully applied to reproduce light curves and spectra of
normal bright SNe~Ia
(Harkness, 1991).

 The second scenario for progenitor
models consists of two white dwarfs in a close orbit which
decays due to the emission of gravitational radiation and this, eventually,
leads to the merging of the two WDs. In an intermediate step,
these models form a low density WD surrounded by a CO envelope
 (Webbink 1984; Iben \& Tutukov 1984; Paczy\'nski 1985, Iben 1997).

A third class of models involves double detonation of a C/O-WD
triggered by detonation of a helium layer in low-mass white dwarfs
as explored by Nomoto (1980),
Woosley, Weaver \& Taam (1980), and most recently by Woosley \&  Weaver (1994),
Livne \&  Arnett (1995)
 and H\"oflich \& Khokhlov (1996).
{}From light curves and spectra, this scenario can be excluded as accounting
for the
majority of SNe~Ia events (H\"oflich et al. 1997, Nugent et al. 1997).

 A separate but closely  related uncertainty is the evolution of the progenitor
systems (e.g.
Canal et al. 1995, Hernanz et al. 1997).
 If the progenitor
population undergoes a change with time, methods to determine  $\Omega_M$ and
$\Lambda $
 (and hence $q_o$)  using only  local calibrators could be systematically
flawed.

 Time evolution is expected to  produce the following main effects: (a) a
lower
  metallicity will
decrease the time scale for  stellar evolution of individual stars
by about 20 \% from Pop I to Pop II stars (Schaller et al. 1992) and,
consequently,
the  progenitor population which contributes to the SNe~Ia rate at any given
time.
 The stellar radius also shrinks. This will influence the statistics
of systems with mass overflow;
 (b) Evolutionary effects of the stellar population will change the
mass function present at the time corresponding to a given redshift;
 (c) The initial metallicity will determine the electron to nucleon fraction of
the outer layers
 and hence affects
the products of nuclear burning;
 (d) Systems with a shorter life time may dominate early on and, consequently,
the typical C/O ratio
 of the central region of the WD will be reduced;
 (e) The properties of the interstellar medium may change;
 (f) In principle, a change of the metallicity may alter the M-R relation. Fig.
1 shows that
 this issue is not a major concern.

 \begfig 0.1cm
\psfig{figure=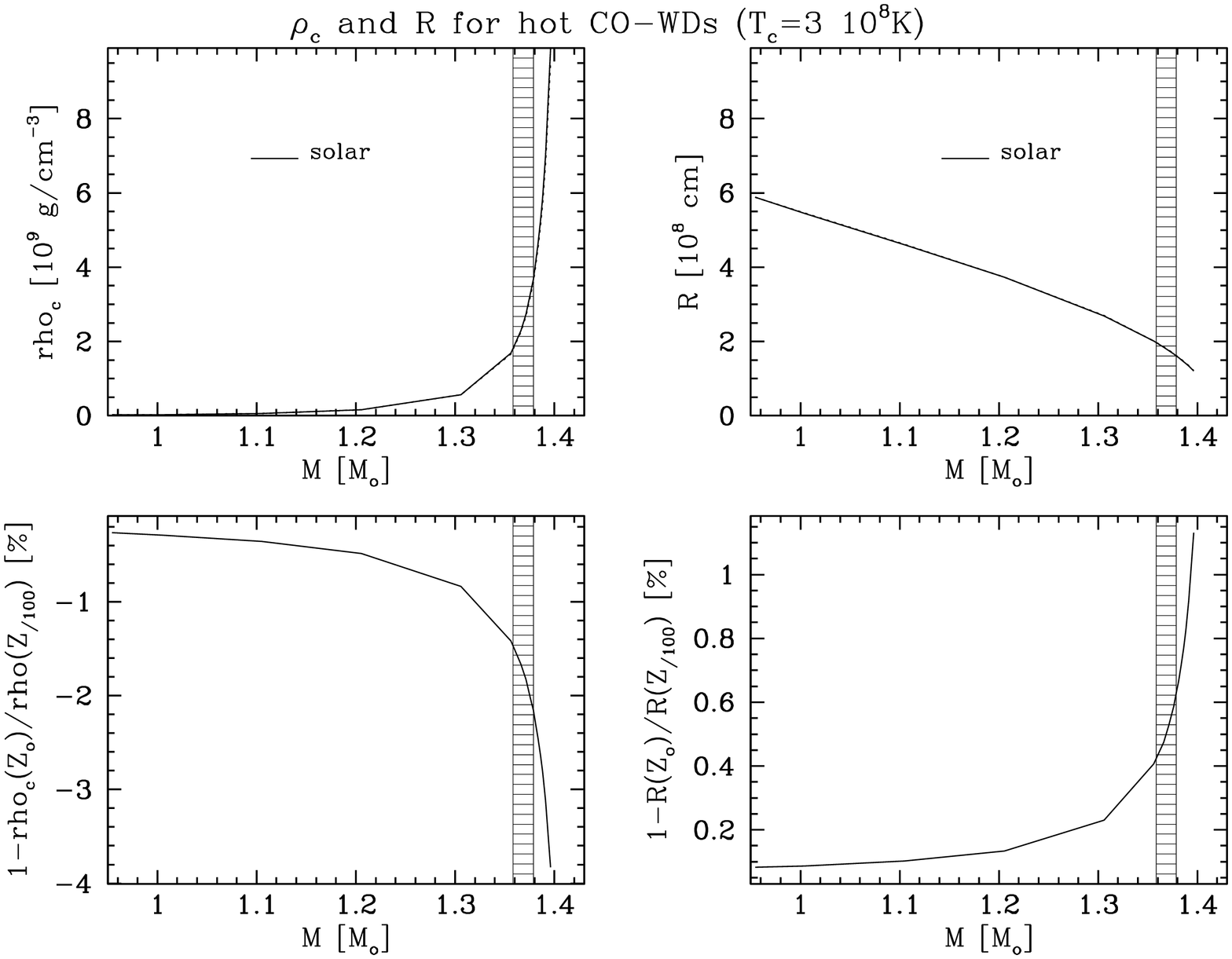,width=15.6cm,rwidth=9.5cm,clip=,angle=270}
\figure{1}{ Influence of the metallicity on the structure of a WD. The shaded band   marks
the region which allows for a successful reproduction of observed LCs and spectra for SNe~Ia
(H\"oflich \& Khokhlov 1996). In the upper plots,
 the M-R relation  and the $\rho_c$-M relations are given
for solar metallicities and, in the lower plots, the difference (in \%) for
models with 1/100 of solar metallicity are given.
 On the scales of the upper plots, the curves corresponding to different
metallicities would merge.}
\endfig

 In this paper we  address the possible influence of evolutionary effects on
light curves and spectra for the delayed detonation scenario.
 Because the time evolution of the composition is not well known and because
we have not considered the entire variety of possible models, the values given
below
{\sl do not } provide a basis  for quantitative corrections of existing
observations.
The goal is to get a first order estimate of the size of the systematic effects
and
to demonstrate how evolutionary effects can be recognized  in a real data
sample and how
 one may be able to correct for them in the determination of  cosmological
parameters.
 In \S 2, the numerical treatment is briefly described. In \S  3,
 the influence of metallicity and the variation of the C/O ratio on
the isotopic abundances,  density and velocity structure of the envelope are
discussed.
 Light curves and maximum light spectra  are presented.
We discuss briefly how the effects  of metallicity can be distinguished from
mixing processes.
In \S 4,
the influence of evolutionary effects on
the  use of SNe~Ia to determine cosmological parameters is presented.
Section 5 gives a  discussion of our results and conclusions.

\section{Brief Description of the Numerical Methods}

\subsection { Hydrodynamics}

 The explosions are calculated using a  one-dimensional
  radiation-hydro code, including nuclear networks  (H\"oflich \& Khokhlov
 1996).
 This code
solves the hydrodynamical equations explicitly
by the piecewise parabolic method (Collela \&  Woodward 1984) and includes the
solution of the radiation transport
implicitly via  moment equations, expansion opacities,
 and a  detailed equation of state (H\"oflich et al. 1993).  About five hundred
depth
points are used.  Radiation transport has been included
to provide a smoother transition from the hydrodynamical explosion to the phase
of free expansion.  We omit $\gamma$-ray transport during the hydrodynamical
phase  because of the
high optical depth of the inner layers.
 Nuclear burning is taken into account using our network which has been tested
in many explosive
environments (see e.g. Thielemann, Nomoto \&   Hashimoto 1996
 and references therein).
 During the hydrodynamical calculations,
 an $\alpha$-network of 14 isotopes is included to
describe the energy release.
The final chemical structure is calculated  by post-processing the
hydrodynamical model using
a network of 216 nuclei.
 The accuracy of the energy release in the reduced network has been found
to be about 1 to 3 \% .

\subsection { Light Curves}

Based on the explosion models, the subsequent expansion
 and  bolometric as well as monochromatic light
curves are calculated using a scheme recently developed, tested
and widely applied to  SN Ia (e.g. H\"oflich et al. 1993, H\"oflich et al. 1996
and references therein).
The code used in this phase is similar to that described above, but
 nuclear burning is neglected and
 $\gamma $ ray transport is included via a Monte Carlo scheme (H\"oflich,
M\"uller \& Khokhlov 1992).
In order to allow for a more consistent treatment of scattering, we
solve both the (two lowest) time-dependent radiation moment equations for the
radiation
energy and the radiation flux, and a total energy equation.
At each time step, we then use $T(r)$ to determine the
Eddington factors and mean opacities by solving the frequency-dependent
radiation transport equation in the comoving frame (see next section) in about
100 frequency bands (see below)
and integrate to obtain the frequency-averaged Eddington factors.  We  use the
latter to iterate the solution with the frequency-integrated
energy and flux equations.
 The frequency averaged  opacities, i.e. Planck, Rosseland and Flux means, have
been calculated under the assumption
of local thermodynamical equilibrium. This is a reasonable approximation
for the light curve  since diffusion time scales are always governed by layers
of large optical depths.
 Note that the comparison of L(r) between the frequency independent
solution and the  frequency dependent solution provides a critical test for the
consistency of
the approximations used in the radiation hydro.

Both the monochromatic and mean opacities are calculated using the Sobolev
approximation
(Sobolev 1957) to
calculate the absorption probability within a shell and to include line
blanketing. The approach
is similar to Karp et al. (1977) but generalized for the comoving frame and the
integration boundaries are
adjusted to a radial grid (H\"oflich 1990).
The scattering, photon redistribution  and thermalization terms
used in the light curve opacity calculation are calibrated with NLTE
calculations
using the formalism of the equivalent-two-level approach
(H\"oflich 1995).
\noindent
\begfig 0.1cm
\hsize=6.0cm
 \psfig{figure=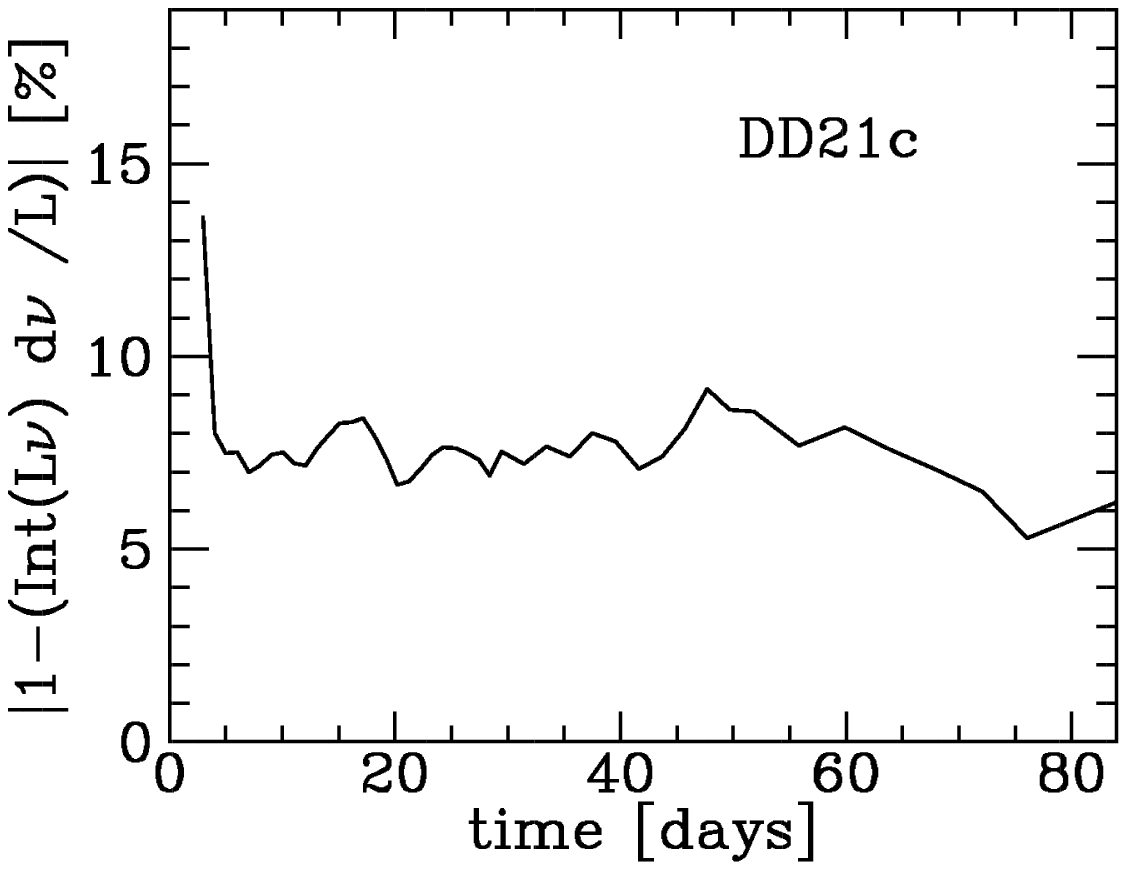,width=15.6cm,rwidth=9.5cm,clip=,angle=270}
\vskip -4.5cm
\figure{2}{Test for the consistency of the integrated luminosity based on
the solution of the monochromatic  and frequency integrated radiation transport equation
for model DD21c (Table 1). About       780 frequencies have been used in the former case.
 Note that this inconsistency only enters the Eddington factors 
which are based on the solution of monochromatic transport equation, but 
not the energy conservation.}
\endfig

To calculate the monochromatic light curves, we use the $T(r)$
with  the time dependence of the structure given by the frequency-integrated
solution of the momentum equations
 to solve  the frequency-dependent transfer equations in LTE every 0.2 to 0.5
days and
  to get $L_\nu$ in the observer's frame.
 The broad band light curves are determine by convolution
of $L_\nu$ with the filter functions.
We use a few hundred
frequency bands with the  scheme  described in the following subsection.
 Note that a proper treatment of the frequency derivatives in the comoving
frame
equations requires the use of about 5 to 10 times more frequencies than
frequency bands.
 To test the consistency
between the frequency-dependent and frequency-independent calculations
we also integrate  the $L_\nu$  and check to see how close the resulting L
is to that from the solution of the frequency-integrated moment equations. The
solutions
are consistent within 10 \% as shown in Fig. 2  (see also Khokhlov et al.
1993).
 Monochromatic colors from our light curve code have been compared to colors
calculated
by our detailed NLTE spectral code (H\"oflich 1995).
 Based on this comparison, the solutions for the Type Ia models are good
to a few percent near maximum light and deteriorate to 0.4 mag at
about 100 days when we stop the LC calculation.

\subsection {Spectra}

Finally, detailed NLTE spectra have been constructed based on
the  LC calculations.
Thus, the effect of energy stored during previous epochs is properly taken into
account.
 The energetics of the model are calculated. Given an explosion model, the
evolution of
 the spectrum is not subject to any tuning or free parameters such as the total
luminosity.
 A modified version of our  code for Nlte Extended ATmospheres (NEAT) is used.
Although time dependence in the rate equations can be included in this code, it
has been omitted because
it was found to be negligible.
 For details  see   H\"oflich (1990, 1995), H\"oflich et al. (1997) and
references therein.

For the NLTE spectra,
 the  density, chemical profiles and the luminosity
as a function of the radial distance $r$ are given by  the hydrodynamical
explosion and  the
 LC calculations, including  the Monte-Carlo scheme for  $\gamma $-ray
  transport.
  The radiation transport equation (including relativistic terms)
 is solved in the comoving frame
 according to Mihalas, Kunacz \& Hummer (1975, 1976,  1982).
 Blocking by weak lines  is included in
   a `quasi' continuum approximation, i.e. the frequency derivative terms in
the radiation
transport equation are included in the narrow line limit  to calculate
the probability for photons  to pass a radial sub-shell along a given direction
$\mu $
(Castor, 1974,  Abbot and Lucy  1985, H\"oflich, 1990).
   The statistical equations are solved consistently  with the radiation
transport
 equation to determine the non-LTE occupation numbers using both an accelerated
lambda
iteration (see  Olson  et al.  1986)    and an  equivalent-two level approach
for
 transitions from the ground state. That
provides an efficient way to take the non-thermal
fraction of the source function into account during the radiation transport
and, effectively,
accelerates both the convergence rate and the stability of the systems
(H\"oflich 1990, 1995).
 A comparison of the
    explicit with the implicit source functions provides a sensitive tool to
test for
convergence of the system of rate and radiation transport equations.

 Excitation by   gamma rays
is included. Detailed atomic models are used for up to the three most abundant
ionization stages
of several elements, i.e. C, O, Ne, Na, Mg, Si, S, Ca, Fe, Co, Ni, taking into
account
20 - 30 levels and  80 - 180 transitions in the main ionization stage. Here, we
use
detailed term-schemes for C~II, O~II, Ne I, Na~I, Mg~II, Si~II, S~II, Ca~II,
Fe~II, Co~II and Ni~II.
 The corresponding lower and upper ions are represented by the ground states.
 The energy levels and cross sections of  bound-bound transitions are taken
from
 Kurucz (1993ab, 1996).
 In addition to $\approx $ 10,000 lines treated in full
 NLTE, a total of $\approx 1,000,000$ lines out of
a list of 31,000,000 (Kurucz 1993a) are included for the radiation transport.
 For these lines, we assume
LTE population numbers for each ion. To calculate the ionization balance,
 excitation by  hard radiation is taken into
 account. LTE-line scattering is not taken as a free parameter (Nugent et al.
1997). Instead,
  LTE-line scattering is  included using  an equivalent-two-level
approach, calibrated by the elements treated in full NLTE.

\section{Results}

\subsection{Explosion Models}

 The influence of the initial metallicity and  mixing on light curves and
spectra
 has been studied for the example of a set of delayed detonation models with
DD21c being
the reference model. In Table 1,
  the quantities given
 in columns 2 to 7 are:  $M_\star$ WD mass; $\rho_c$ central
 density of the WD (in $10^9 g~cm^{-3}$), $\alpha$ ratio of the deflagration
velocity and
 local sound speed; $\rho_{tr}$ transition density (in $10^7 g~cm^{-3}$) at
which the
 deflagration is assumed to turn into a detonation; C/O ratio;     $R_Z$ the
metallicity relative to solar by mass; $E_{kin}$ kinetic energy (in
$10^{51}erg$);
$M_{Ni}$ mass of $^{56}Ni$ (in solar units).
The parameters  are close to those which
reproduce both the spectra and light curves reasonably well (Nomoto et al.
1984; H\"oflich
1995; H\"oflich \& Khokhlov 1996).

\begtab
\table{1}
{ Basic parameters for 
the  delayed detonation models.
}
 
\hline 
\+Model~~~~~ &  $M_\star$ ~~~~~& $\rho _c$ ~~~~~~~~~ & $\alpha$ ~~~~~&   $\rho_{tr}$~~~~~~~~~~~~ & ~~C/O~~ & ~~$R_Z$~~ & $E_{kin}$~~~~~  & $M_{Ni}$~~~~~  \cr
%
\hline                     
\+DD13c   & 1.4  &  2.6   &  0.03  &  3.0 & 1/1 & 1/1 & 1.36 &
                           0.79  \cr
\+DD21c   &  1.4  &  2.6   &  0.03  &  2.7 & 1/1 & 1/1  &  1.32  &
                           0.69  \cr
\+DD23c   &  1.4  &  2.6   &  0.03  &  2.7 & 2/3 & 1/1 &  1.18  &
                           0.59  \cr
\+DD24c   &  1.4  &  2.6   &  0.03  &  2.7 & 1/1 & 1/3  &  1.32  &
                           0.70  \cr
\+DD25c   &  1.4  &  2.6   &  0.03  &  2.7 & 1/1 & 3/1  &  1.32  &
                           0.69  \cr
\+DD26c   &  1.4  &  2.6   &  0.03  &  2.7 & 1/1 & 1/10  &  1.32  &
                           0.73  \cr
\+DD27c   &  1.4  &  2.6   &  0.03  &  2.7 & 1/1 & 10/1  &  1.32  &
                           0.69  \cr
\hline         
\endtab

\begfig -.4cm
 \psfig{figure=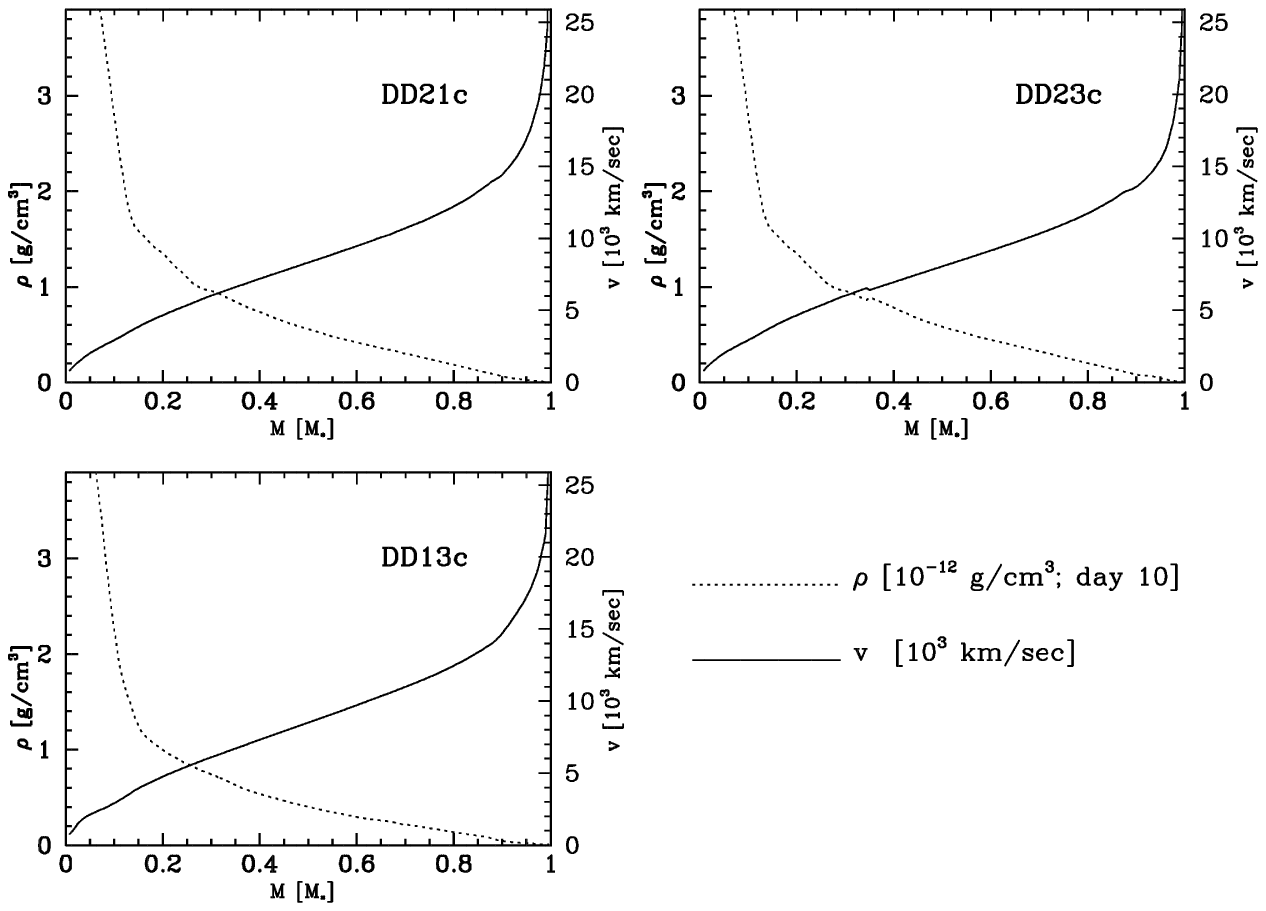,width=15.0cm,rwidth=9.0cm,clip=,angle=270}
\figure{3}{ Density and velocity 
as a function of  mass for three delayed detonation models (Table 1).\hfill}
\endfig

\begfig 0.1cm
 \psfig{figure=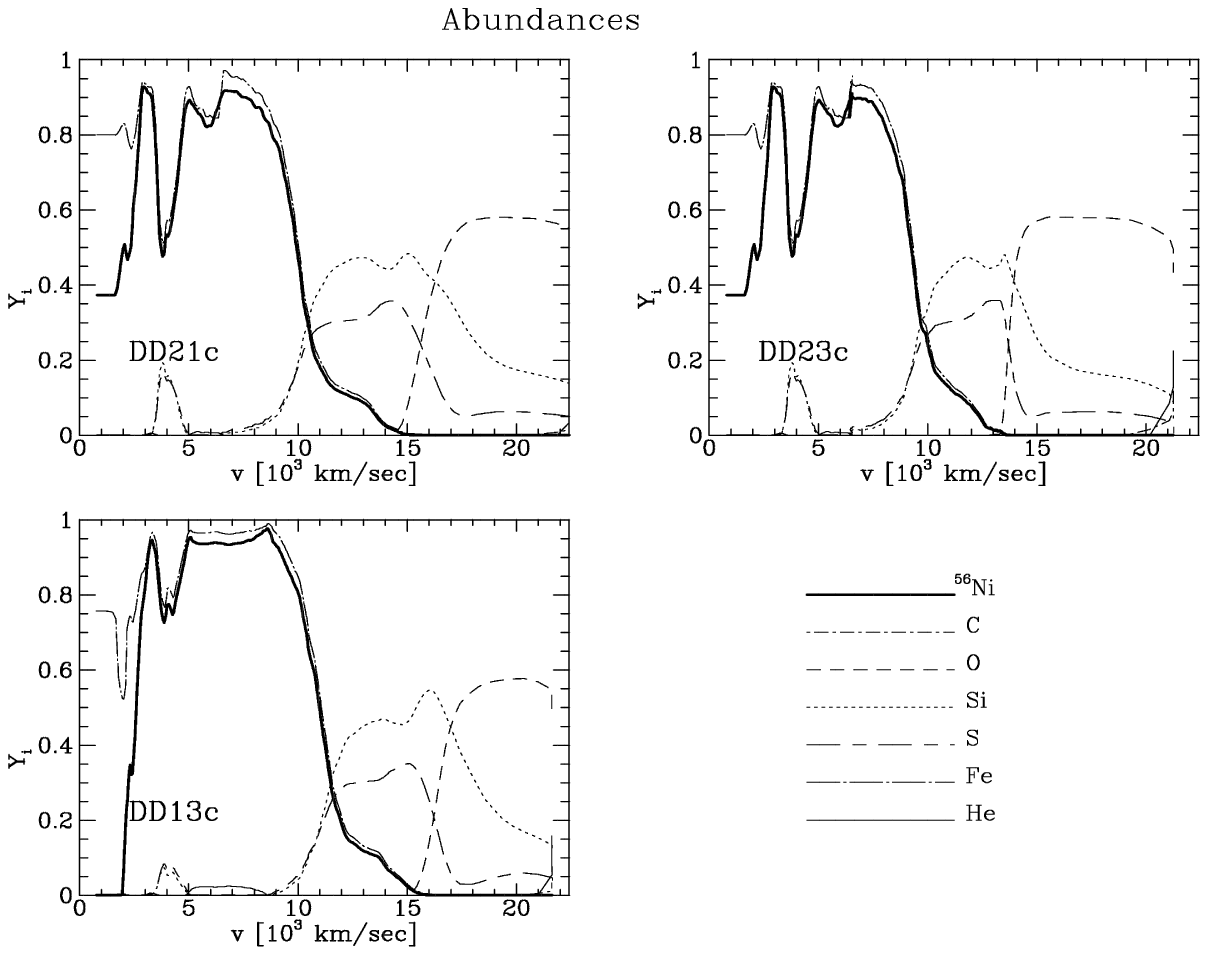,width=15.0cm,rwidth=9.0cm,clip=,angle=270}
\figure{4}{ Abundances as a function of the final expansion velocity for the
 three delayed detonation models of Figure 3. 
Both the initial $^{56}Ni$ and the final Fe profiles are shown.}
\endfig

\begtabsmall    
\table{2}{Total abundances of model DD21c with solar  initial composition.
}
\hline
 \+ ~~~~He ~~~~~~ & ~~~~~ C~~~~~~ & ~~~~~ O~~~~~~ & ~~~~~Ne~~~~~~ & ~~~~~Na~~~~~~ & ~~~~~Mg~~~~~~ & ~~~~~Al~~~~~~ & \cr
 \hline
 \+   1.16E-03 &   5.44E-04 &   5.03E-02 &   1.67E-03 &   1.05E-05 &   1.02E-02 &   6.12E-05 & \cr
 \hline
 \+ ~~~~~Si~~~~~~ & ~~~~~ P~~~~~~ & ~~~~~ S~~~~~~ & ~~~~~Cl~~~~~~ & ~~~~~Ar~~~~~~ & ~~~~~ K~~~~~~ & ~~~~~Ca~~~~~~ & \cr
 \hline
 \+   2.06E-01 &   2.30E-05 &   1.49E-01 &   9.50E-06 &   3.56E-02 &   7.50E-06 &   4.00E-02 & \cr
 \hline
 \+ ~~~~~ V~~~~~~ & ~~~~~Cr~~~~~~ & ~~~~~Mn~~~~~~ & ~~~~~Fe~~~~~~ & ~~~~~Co~~~~~~ & ~~~~~Ni~~~~~~ & ~~~~~Cu~~~~~~ & \cr
 \hline
 \+   1.02E-03 &   2.20E-02 &   1.44E-02 &   6.87E-01 &   1.28E-02 &   2.19E-02 &   1.61E-05 & \cr
 \hline

\endtab         

\begfig -.4cm
 \psfig{figure=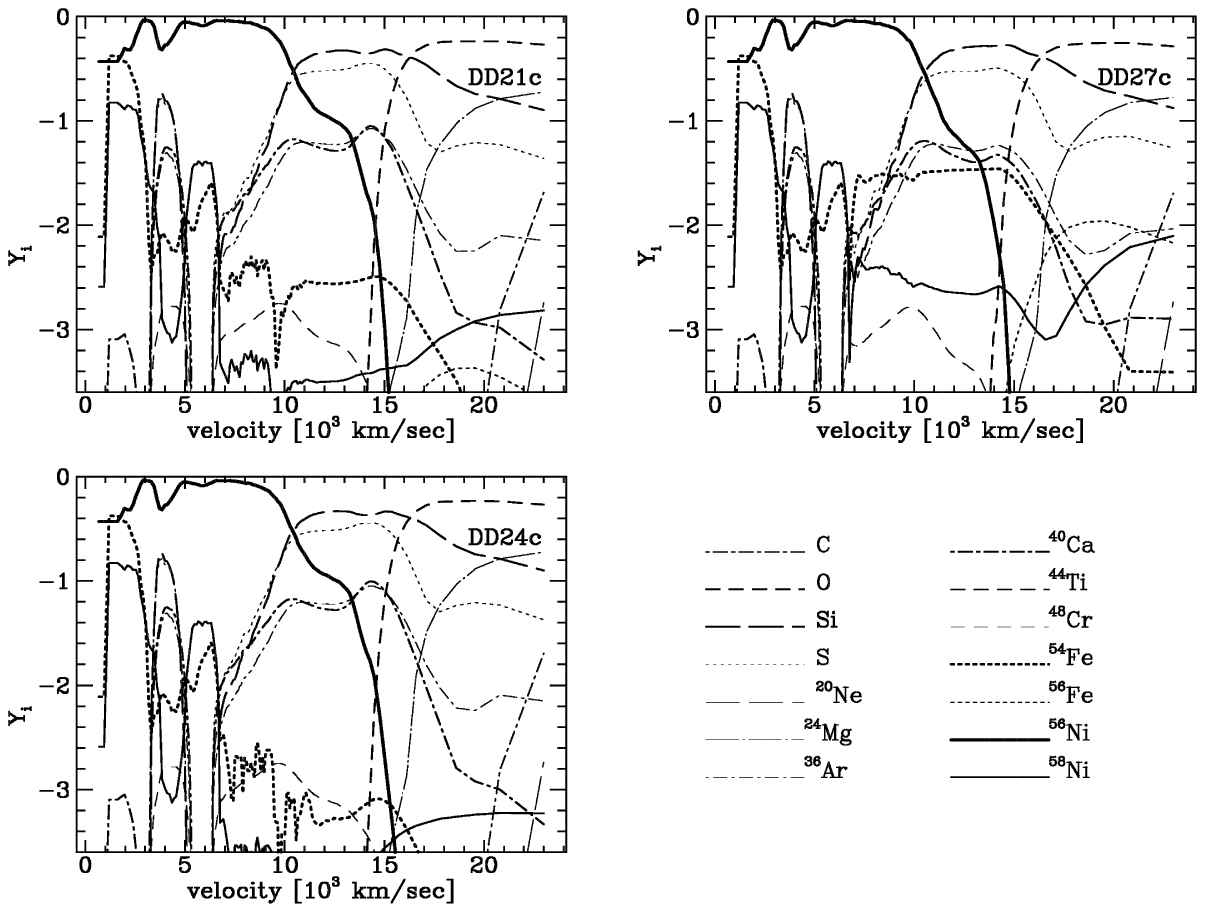,width=15.0cm,rwidth=9.0cm,clip=,angle=270}
\figure{5}{Abundances of different isotopes as a function of the
expansion velocity for models 
DD21c, DD24c and DD27c that have initial compositions of 
solar, 1/3 solar, and 1/10 of solar, respectively.}
\endfig

\begtabsmall    
\table{3}{Isotopes  $M_{i}$ in $M_\odot$  with $M_{i}    \geq 1.E-10$
 for delayed detonation models (see Table 1).
}
\hline
\+ Isotope ~~~~ & ~~DD21c~~~~~ &  ~~DD13c~~~~~ & ~~DD23c~~~~~ & ~~DD24c~~~~~ & ~~DD25c~~~~~ &
 ~~DD26c~~~~~ &  ~~DD27c~~~~~ \cr
\hline 
 \+$^{ 4}$He &   1.16E-03 &   7.33E-03 &   2.58E-03 &   1.16E-03 &   1.16E-03 &   1.16E-03 &   1.16E-03 \cr
 \+$^{12}$ C &   5.44E-04 &   3.50E-04 &   1.04E-03 &   5.38E-04 &   5.39E-04 &   5.40E-04 &   5.33E-04 \cr
 \+$^{15}$ N &   6.52E-09 &   7.20E-09 &   8.09E-09 &   1.78E-08 &   2.18E-09 &   4.66E-08 &   5.38E-10 \cr
 \+$^{16}$ O &   5.03E-02 &   4.65E-02 &   6.80E-02 &   4.98E-02 &   5.01E-02 &   4.99E-02 &   4.88E-02 \cr
 \+$^{20}$Ne &   1.67E-03 &   1.54E-03 &   2.06E-03 &   1.67E-03 &   1.65E-03 &   1.68E-03 &   1.61E-03 \cr
 \+$^{21}$Ne &   1.27E-08 &   6.35E-09 &   2.22E-08 &   1.16E-08 &   1.43E-08 &   1.12E-08 &   1.82E-08 \cr
 \+$^{22}$Na &   2.21E-08 &   1.13E-08 &   4.18E-08 &   2.39E-08 &   1.86E-08 &   2.47E-08 &   1.56E-08 \cr
 \+$^{23}$Na &   1.05E-05 &   7.68E-06 &   1.36E-05 &   1.02E-05 &   1.12E-05 &   1.01E-05 &   1.22E-05 \cr
 \+$^{24}$Mg &   1.02E-02 &   1.09E-02 &   1.54E-02 &   1.02E-02 &   1.00E-02 &   9.52E-03 &   9.36E-03 \cr
 \+$^{25}$Mg &   8.09E-07 &   8.47E-07 &   1.59E-06 &   2.77E-07 &   1.20E-06 &   1.57E-07 &   1.95E-06 \cr
 \+$^{26}$Mg &   6.26E-07 &   5.01E-07 &   8.89E-07 &   4.18E-07 &   1.23E-06 &   3.17E-07 &   3.21E-06 \cr
 \+$^{27}$Al &   6.12E-05 &   6.62E-05 &   7.68E-05 &   4.90E-05 &   1.01E-04 &   5.03E-05 &   1.68E-04 \cr
 \+$^{28}$Si &   2.06E-01 &   1.65E-01 &   2.12E-01 &   2.07E-01 &   2.09E-01 &   2.08E-01 &   2.11E-01 \cr
 \+$^{29}$Si &   7.67E-05 &   8.40E-05 &   1.17E-04 &   4.84E-05 &   1.03E-04 &   3.15E-05 &   1.43E-04 \cr
 \+$^{30}$Si &   3.10E-05 &   1.83E-05 &   1.83E-05 &   7.69E-05 &   2.12E-05 &   1.11E-04 &   7.63E-05 \cr
 \+$^{31}$ P &   2.30E-05 &   1.79E-05 &   2.74E-05 &   3.71E-05 &   2.41E-05 &   6.01E-05 &   4.74E-05 \cr
 \+$^{32}$ S &   1.49E-01 &   1.12E-01 &   1.53E-01 &   1.53E-01 &   1.46E-01 &   1.56E-01 &   1.39E-01 \cr
 \+$^{33}$ S &   5.25E-05 &   3.79E-05 &   7.45E-05 &   3.57E-05 &   6.91E-05 &   2.53E-05 &   1.03E-04 \cr
 \+$^{34}$ S &   2.19E-05 &   2.03E-05 &   1.83E-05 &   2.66E-05 &   8.66E-05 &   3.96E-05 &   3.43E-04 \cr
 \+$^{35}$Cl &   9.50E-06 &   6.33E-06 &   1.60E-05 &   1.17E-05 &   1.45E-05 &   1.80E-05 &   3.14E-05 \cr
 \+$^{36}$Ar &   3.56E-02 &   2.60E-02 &   3.74E-02 &   3.70E-02 &   3.41E-02 &   3.80E-02 &   3.11E-02 \cr
 \+$^{37}$Cl &   7.75E-06 &   4.31E-06 &   1.22E-05 &   5.63E-06 &   9.62E-06 &   3.94E-06 &   1.44E-05 \cr
 \+$^{38}$Ar &   1.28E-05 &   1.09E-05 &   1.23E-05 &   8.52E-07 &   6.95E-05 &   1.03E-06 &   2.64E-04 \cr
 \+$^{39}$ K &   7.50E-06 &   4.24E-06 &   2.34E-05 &   3.31E-06 &   1.54E-05 &   3.82E-06 &   3.05E-05 \cr
 \+$^{40}$Ca &   4.00E-02 &   2.94E-02 &   4.23E-02 &   4.19E-02 &   3.77E-02 &   4.30E-02 &   3.34E-02 \cr
 \+$^{41}$Ca &   1.87E-06 &   8.10E-07 &   5.84E-06 &   1.31E-06 &   2.43E-06 &   9.25E-07 &   3.90E-06 \cr
 \+$^{42}$Ca &   2.95E-07 &   1.89E-07 &   6.00E-07 &   8.80E-09 &   1.63E-06 &   9.36E-09 &   6.71E-06 \cr
 \+$^{43}$Ca &   6.33E-09 &   5.62E-08 &   4.59E-07 &   9.07E-09 &   6.47E-09 &   1.84E-08 &   1.41E-08 \cr
 \+$^{44}$Ca &   3.04E-05 &   3.65E-05 &   8.38E-05 &   3.13E-05 &   2.92E-05 &   3.18E-05 &   2.73E-05 \cr
 \+$^{45}$Sc &   1.56E-07 &   7.57E-08 &   3.08E-06 &   1.42E-07 &   1.84E-07 &   1.18E-07 &   2.46E-07 \cr
 \+$^{46}$Ca &   2.60E-09 &   2.36E-09 &   4.21E-09 &   6.14E-10 &   8.54E-09 &   1.75E-10 &   2.11E-08 \cr
 \+$^{48}$Ca &   9.43E-09 &   8.07E-09 &   1.44E-08 &   3.15E-09 &   2.76E-08 &   9.59E-10 &   8.74E-08 \cr
 \+$^{46}$Ti &   3.03E-07 &   4.15E-06 &   8.37E-06 &   1.37E-07 &   1.00E-06 &   1.24E-07 &   4.15E-06 \cr
 \+$^{47}$Ti &   4.55E-07 &   3.94E-06 &   1.96E-05 &   4.41E-07 &   4.99E-07 &   4.49E-07 &   6.62E-07 \cr
 \+$^{48}$ V &   1.02E-03 &   7.92E-04 &   1.63E-03 &   1.03E-03 &   9.80E-04 &   1.04E-03 &   8.92E-04 \cr
 \+$^{49}$ V &   2.75E-05 &   1.92E-05 &   2.04E-04 &   2.37E-05 &   3.46E-05 &   1.87E-05 &   4.62E-05 \cr
 \+$^{50}$Cr &   8.73E-05 &   9.16E-05 &   6.91E-04 &   7.69E-05 &   1.27E-04 &   7.49E-05 &   2.70E-04 \cr
 \+$^{51}$Cr &   1.23E-04 &   3.17E-05 &   1.38E-03 &   1.16E-04 &   1.39E-04 &   1.16E-04 &   1.70E-04 \cr
 \+$^{52}$Cr &   2.20E-02 &   1.57E-02 &   2.83E-02 &   2.21E-02 &   2.15E-02 &   2.20E-02 &   2.02E-02 \cr
 \+$^{50}$Mn &   6.96E-07 &   5.64E-07 &   1.49E-05 &   6.97E-07 &   6.95E-07 &   7.03E-07 &   6.94E-07 \cr
 \+$^{53}$Mn &   1.07E-03 &   7.17E-04 &   4.25E-03 &   9.71E-04 &   1.26E-03 &   9.16E-04 &   1.60E-03 \cr
 \+$^{54}$Mn &   3.42E-09 &   6.65E-07 &   7.82E-07 &   1.68E-09 &   2.17E-08 &   1.63E-09 &   1.49E-07 \cr
 \+$^{55}$Mn &   1.44E-02 &   4.21E-03 &   3.41E-02 &   1.40E-02 &   1.54E-02 &   1.38E-02 &   1.72E-02 \cr
 \+$^{54}$Fe &   4.17E-02 &   4.84E-02 &   6.83E-02 &   4.04E-02 &   4.65E-02 &   4.00E-02 &   6.22E-02 \cr
 \+$^{56}$Fe &   6.87E-01 &   7.75E-01 &   5.63E-01 &   6.89E-01 &   6.83E-01 &   6.97E-01 &   6.79E-01 \cr
 \+$^{57}$Co &   1.28E-02 &   9.43E-03 &   1.83E-02 &   1.27E-02 &   1.33E-02 &   1.26E-02 &   1.45E-02 \cr
 \+$^{58}$Ni &   2.19E-02 &   2.58E-02 &   3.05E-02 &   2.18E-02 &   2.24E-02 &   2.17E-02 &   2.41E-02 \cr
 \+$^{59}$Ni &   1.34E-04 &   9.92E-04 &   2.95E-04 &   1.29E-04 &   1.46E-04 &   1.28E-04 &   1.82E-04 \cr
 \+$^{60}$Cu &   1.61E-05 &   9.04E-03 &   1.65E-05 &   1.63E-05 &   1.60E-05 &   1.67E-05 &   1.60E-05 \cr
 \+$^{61}$Cu &   1.95E-06 &   3.85E-04 &   4.91E-06 &   1.93E-06 &   1.99E-06 &   1.92E-06 &   2.04E-06 \cr
 \+$^{62}$Cu &   4.17E-06 &   3.12E-04 &   3.70E-06 &   4.17E-06 &   4.17E-06 &   4.16E-06 &   4.17E-06 \cr
 \+$^{63}$Zn &   3.66E-08 &   1.66E-03 &   9.42E-08 &   3.42E-08 &   3.61E-08 &   3.17E-08 &   3.33E-08 \cr
 \+$^{64}$Zn &   9.62E-08 &   5.25E-03 &   2.10E-07 &   3.15E-08 &   2.69E-07 &   1.23E-08 &   8.54E-07 \cr
 \+$^{65}$Zn &   1.49E-08 &   2.28E-05 &   1.85E-08 &   4.27E-09 &   4.61E-08 &   1.27E-09 &   1.33E-07 \cr
 \+$^{66}$Ga &   2.16E-09 &   6.81E-06 &   2.94E-09 &   1.16E-09 &   3.83E-09 &   7.72E-10 &   5.68E-09 \cr
 \+$^{67}$Ge &   1.55E-09 &   4.96E-06 &   1.85E-09 &   7.79E-10 &   1.86E-09 &   2.65E-10 &   1.17E-09 \cr

\hline
\endtab

 To study the influence of a variation of the C/O ratio,
a model (DD23c) with the same parameters as DD21c has been constructed but with
C/O=2/3
by abundance.
 The total abundances of the most important elements are given for the
reference model DD21c in
Table 2. Fig. 3 gives the density and velocity versus mass for the reference
model DD21c, a model
with a higher transition density (DD13c) and a model with reduced C/O ratio
(DD23c). Fig. 4 gives
the composition profiles of the major elements for the same three models.
 The overall density and velocity structures are insensitive to the transition
density and to the
 C/O ratio.
Qualitatively, the final burning products can be understood by the relation
between
hydrodynamical and the individual nuclear time scales.
 The hydrodynamical time scale is given by
the total energy release during burning of the progenitor. Since most of the
energy
is released by the explosive burning of carbon and oxygen, the energy
production per gram
depends on the initial chemical composition of the WD, i.e. the C/O ratio.
          The nuclear time
scales are determined by the peak temperature during burning, which depends on
the
energy release per volume because the energy density is radiation dominated.
As the energy release per gram is fixed, the peak temperature is given by the
composition
and the local density.
 Thus, the latter two are the dominant factors for the final composition of a
zone.
With decreasing
transition density, less \ni is produced and the intermediate
mass elements expand at lower velocities
 because the later  transition to  detonation allows for a longer
pre-expansion of the  outer layers (DD21c vs. DD13c). Similarly,
with decreasing C/O ratio in the progenitor, the specific energy
release during the  nuclear burning is reduced (DD23c vs. DD21c) and
the transition density is  reached later in time,
resulting in a larger pre-expansion of the outer layers and a more
narrow  region dominated by Si. Moreover, a change of the C/O ratio
from 1 to 2/3 reduces the \ni production by about 20 \%.
 The kinetic energy is reduced by 10 \% corresponding to a 5 \% change in
   the mean expansion velocity.
To first approximation a reduction
 in the transition density
has a similar effect than a decrease
of the C/O ratio or a reduction of the central density (H\"oflich \& Khokhlov
1996).

 To test the influence of the metallicity for $Z\geq 20$ (i.e. nuclei beyond
Ca) we
have constructed models
with parameters identical to DD21c but with initial metallicities
between 0.1 and 10 times solar (Table 1). The energy release,
the density and velocity  structure are virtually identical to that of DD21c.
The main difference is a slight increase of the \ni mass with decreasing
metallicity due to a higher
$Y_e$.
 The reason is that the metallicity mainly effects the initial CNO abundances
of a star.
 These are converted
during the pre-explosion stellar evolution to $^{14}$N in H-burning and via
$^{14}$N($\alpha,\gamma$)$^{18}$F($\beta ^+$)$^{18}$O($\alpha,\gamma$)$^{22}$Ne
to nuclei with
N=Z+2 in He-burning. The result is that increasing metallicity yields
 i.e. a smaller proton to nucleon ratio $Y_e$
throughout the pre-explosive WD (Thielemann et al. 1997). Higher metallicity
and
    smaller $Y_e$
lead to the production of more neutron-rich Fe group nuclei and less $^{56}Ni$.
 For lower metallicity and, thus,
 higher $Y_e$, some additional \ni
is produced  at the expense of $^{54}Fe$  and $^{58}Ni$
  (Thielemann, Nomoto \& Yokoi, 1986).
 The temperature in the inner layers is sufficiently
high during the explosion that
 electron capture  determines $Y_e$. In those layers,
the initial metallicity has no influence on the final burning product.
 The total production of isotopes of all the models is given in Table 3. Fig. 5
gives
the distribution of various isotopes in models DD21c, DD24c,              and
DD27c  which
represent solar, 1/3, and 10 times solar metallicities, respectively.
The main differences due to changes in Z
are in regions with expansion velocities in excess of $\approx $
12000 km/sec.
 Most remarkable is the change in the $^{54}$Fe production which is the
dominant contributor to the abundance  of iron group elements at these
velocities
 since little cobalt has yet been decayed near maximum light.
 For increasing metallicity,
the production of this              isotope increases significantly. For
1/3 solar metallicity, hardly any $^{54}$ Fe  is produced at high velocities,
but $^{54}$Fe is  as high as 5 \% by mass fraction if we start with ten times
the
solar metallicity.
 We note that these layers with v$\geq 12,000~km~ s^{-1}$
 dominate the spectra around maximum light and lines of iron group
elements are an important contributor to the line opacities (see below).

\subsection{Light Curves}

As discussed in previous papers (e.g. H\"oflich et al. 1997 and references
therein),
bolometric and monochromatic light curves provide a valuable tool to probe the
underlying
explosion models, namely the absolute amount of \ni and its distribution.

 \begfig 0.1cm
\hsize=5.5cm
 \psfig{figure=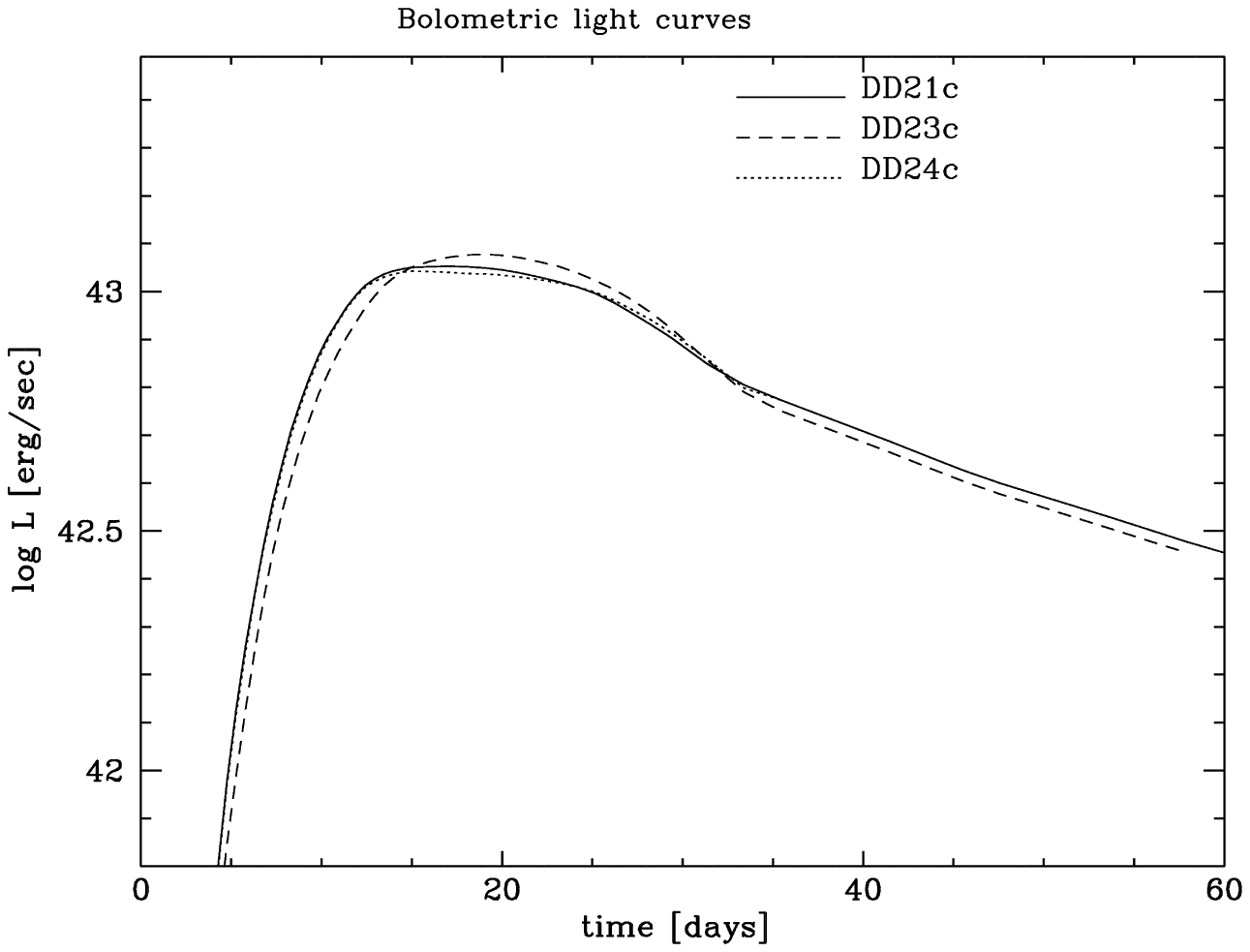,width=14.0cm,rwidth=6.0cm,clip=,angle=270}
\vskip -3.5cm
\figure{6}{Comparison of bolometric light curves of the delayed detonation
models DD21c, DD23c and DD24c with otherwise identical parameters but with different
C/O ratios and metallicity relative to solar (C/O; $R_Z$) of (1;1), (2/3;1) and (1;0.3), 
respectively.
}
\endfig

\begfig 0.1cm
 \psfig{figure=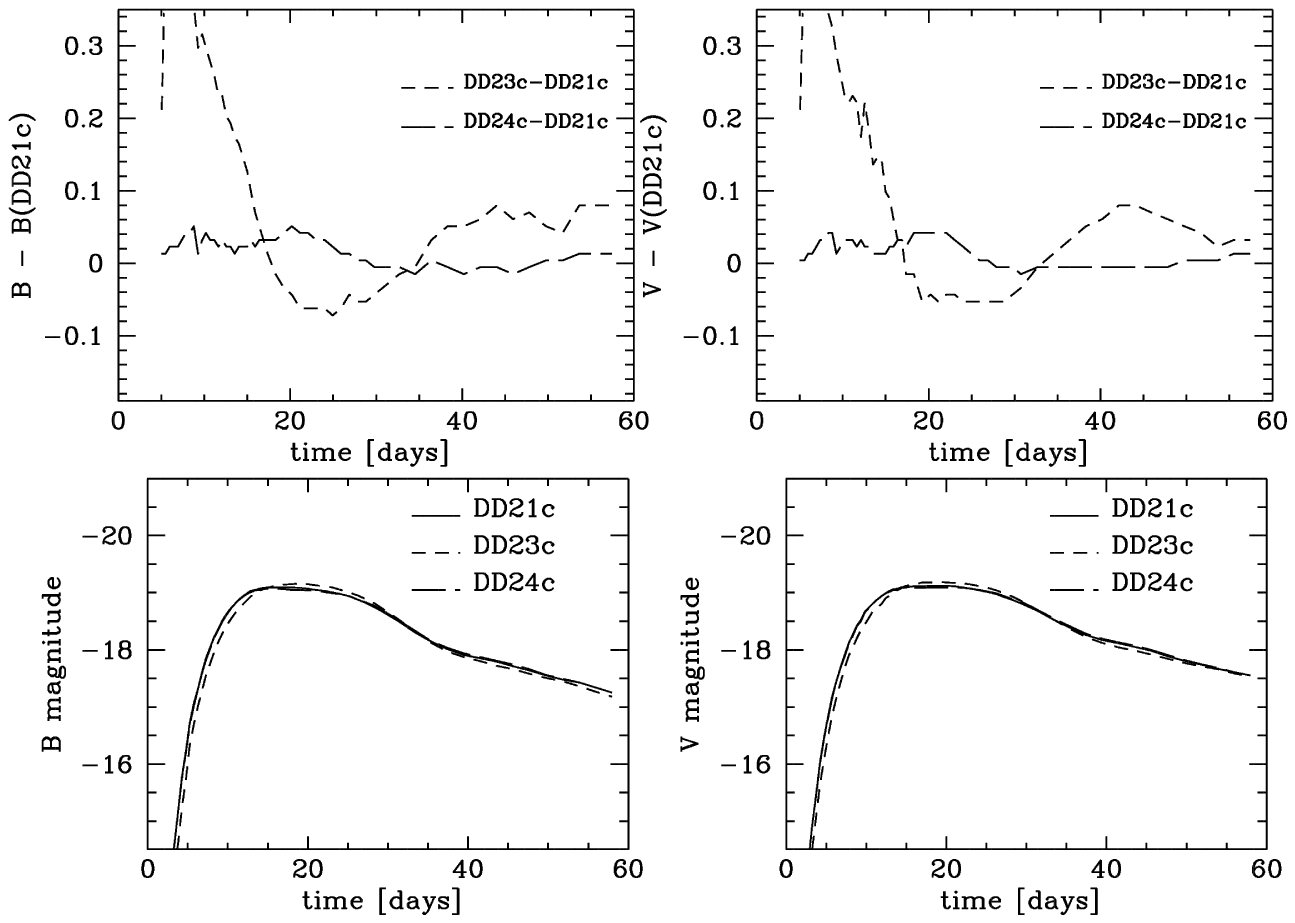,width=15.6cm,rwidth=9.5cm,clip=,angle=270}
\figure{7}{Comparison of  light curves in B (left) and V (right) of the delayed detonation
models DD21c, DD23c and DD24c with otherwise identical parameters but different different
C/O ratios and metallicities relative to solar (C/O; $R_Z$) of (1;1), (2/3;1) and (1;0.3). 
 The difference relative to DD21c and the monochromatic LCs are given
in the upper and lower plots, respectively.
}
\endfig

 Bolometric and monochromatic LCs are shown in Figs. 6 and 7.
 Reducing the ratio  C/O from 1/1 to 2/3 in the WD reduces the \ni production
and the kinetic energy in DD23c compared to DD21c. Both the  bolometric and
monochromatic light
curves are affected.
 The smaller expansion due to the smaller $E_{kin}$ causes a reduced
geometrical dilution of the
matter and a reduction of the expansion work at a given time.
 Consequently, for reduced C/O, the rise time
to maximum light is slower by about 3 days and the maximum brightness is larger
because more of the
stored energy goes into the radiation rather than kinetic energy.
The reduced heating at the later time of maximum light also implies that DD23c
is slightly
redder ($\Delta (B-V) \approx 0.02^m$) than DD21c at its peak.
 After about day 35, however, the
 luminosity  is smaller by about 10 \% for lower C/O
due to the smaller \ni production (i.e. the instantaneous energy production).
The
postmaximum decline is steeper.
The change of the C/O ratio
from 1/1 to 2/3 has a similar effect
on the colors, light curve shape and the distribution of elements
as a 10 \% reduction of the transition density or of the
 central density in delayed detonations.
 For a given ratio of
peak  to tail, a model produced by a reduced C/O ratio will be brighter than a
one obtained by varying the transition density (H\"oflich 1995).
 The  rise time and the   expansion rate as measured by the doppler shift of
lines
(see below)
  provide a way to determine   the C/O ratio and the transition density
independently
 (Table 2, compare  Khokhlov  et al., 1993, H\"oflich 1995, H\"oflich \&
Khokhlov 1996).
The current state of the art for LC calculations and for the progenitor
evolution
does not allow for a sufficiently fine discrimination to determine the absolute
C/O ratio.
A careful
differential analyses of observations may yield information on the relative
variation
if the C/O ratio, and the central ignition density or transition density.

 Changing the initial metallicity Z has very little influence on the bolometric
light curve
(compare DD21c, DD24 in Fig. 6) because
the  \ni  production and energy release vary   by only 4 \% over the entire
range of models.
 In addition, diffusion time scales are mainly determined by deeper layers
which are unaffected by Z.

 Monochromatic LCs are slightly more affected by changes in the initial
metallicity Z (Fig. 7, upper
panel). The principle effect is that
 the radiative cooling in the outer layers increases
with Z.
Consequently, depending on the phase, the B-V is larger
are redder by about 0.02 to 0.04$^m$.
 The absolute  brightness varies by about the same amount.
 The time of  maximum light shifts by only
 $\approx 1$ day.

\subsection {Spectra at Maximum Light}

 Variations in the pattern of the most abundant
elements is similar for changes in the C/O ratio,
 the transition density and the central density of the WD (see Figs. 3 and 4).
 Consequently, the variation of the spectra as a result of changing these
parameters
is similar. Differences would show up in correlation between light curve shape
and
Doppler shifts of lines at a given phase (see above).

\begfig 0.1cm
 \psfig{figure=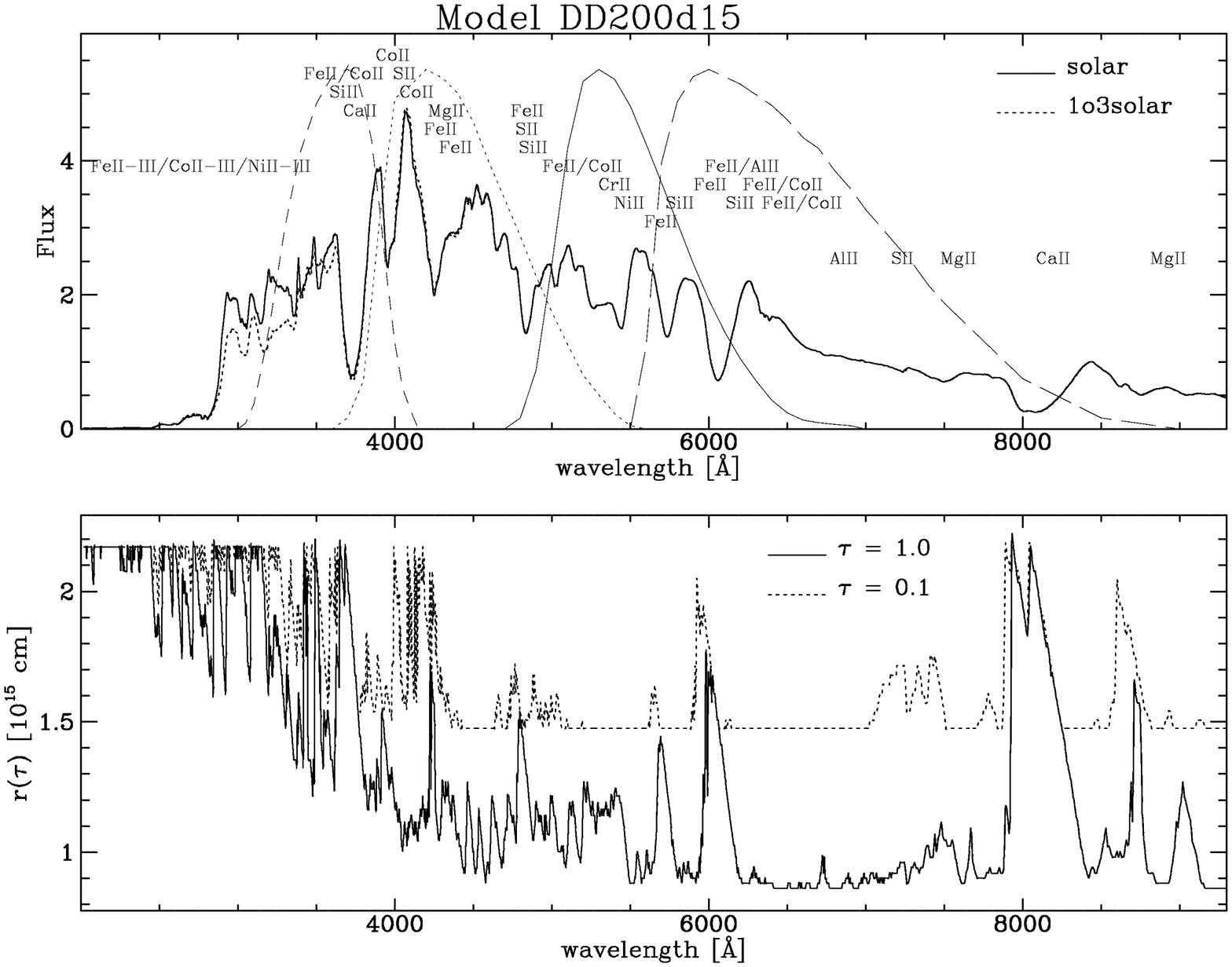,width=15.6cm,rwidth=9.5cm,clip=,angle=270} 
\figure{8}{Comparison of synthetic NLTE spectra at maximum light for
 initial compositions of solar and 1/3 of solar, respectively
(upper graph). The standard Johnson filter functions for UBV, and R are
also shown. In the lower graph, the radius as a function of 
wavelength  where the monochromatic optical depth reaches
0.1 and 1, respectively  is given for DD21c.}
\endfig

 The influence of the initial metallicity Z is interesting primarily
 because it influences the
iron group elements.
As an example, the spectra  of the delayed detonation models with solar and 1/3
solar
metallicity are given in Fig. 8. At maximum light ($\approx $ 17 days), the
line forming region
($\tau \approx 0.1 ~...~1.$) extends
between 1  and 2x$10^{15}$ cm in the optical (Fig. 8, lower graph),
 corresponding to expansion velocities between 8000 and
16,000 $km~s^{-1}$.
In this velocity range the material has been subjected to at least
partial oxygen burning to silicon and the deeper layers are rich
in $^{56}$Ni. This means that the iron peak elements visible in
the optical spectrum near maximum light are freshly synthesized
and do not directly represent the initial metallicity.  The initial
metallicity does affect the optical spectrum near maximum, but
only indirectly and only in the blue.  The abundances of the
intermediate mass elements are not affected by the metallicity,
but the abundance of $^{54}$Fe is a sensitive function of the initial
Z and   $^{54}$Fe provides an important  source of Fe lines to
the opacity for the matter with v $\gta$ 10,000 km s$^{-1}$.  The
iron line opacity is most substantial for wavelengths less than
$\sim$ 4000 \AA.  The effect of the initial metallicity on the spectrum
can be seen in the wavelength region $\sim$ 3000 \AA\ in Fig. 8 (top).
The initial metallicity might be directly observed by examination
of UV spectra that form in the outermost layers with v $\gta$
16,000 to 17,000 km s$^{-1}$ that have not been subject to oxygen burning.

By 2 to 3 weeks after maximum, the spectra are completely insensitive
to the initial Z because the spectrum is formed in even deeper
layers where none of the important abundances are
affected by the metallicity.
 Thus, for two similar bright SNe with similar
expansion velocities, a comparison between the spectral evolution can provide
a method to determine the metallicity difference or may be used to detect
evolutionary effects
for distant SNe~Ia (see below).

 Note that the UV may provide a unique tool to probe the very outer layers
which may have
undergone little burning during the explosion.

\subsection{Mixing during the Explosion}

 Up to now, we have omitted a potential difficulty involved in the
determination of  Z.
As already mentioned,  we may see strong iron group elements in the line
forming region
near maximum
because they are produced at high velocities.
 Alternatively,
these elements may be mixed out during the explosion but after burning took
place.
 To test this,
we have  artificially mixed the composition
 but kept the overall structure fixed
 (see Fig. 9). Unlike a change in the initial metallicity,  mixing alters
the entire spectrum. All burning products will be enhanced in the
outer region including Ni, Co, and Ca. In particular, the Ca lines at about
4000 \AA~ and
in the IR triplet provide a  way to distinguish mixing from metallicity
effects.

\begfig 0.1cm
 \psfig{figure=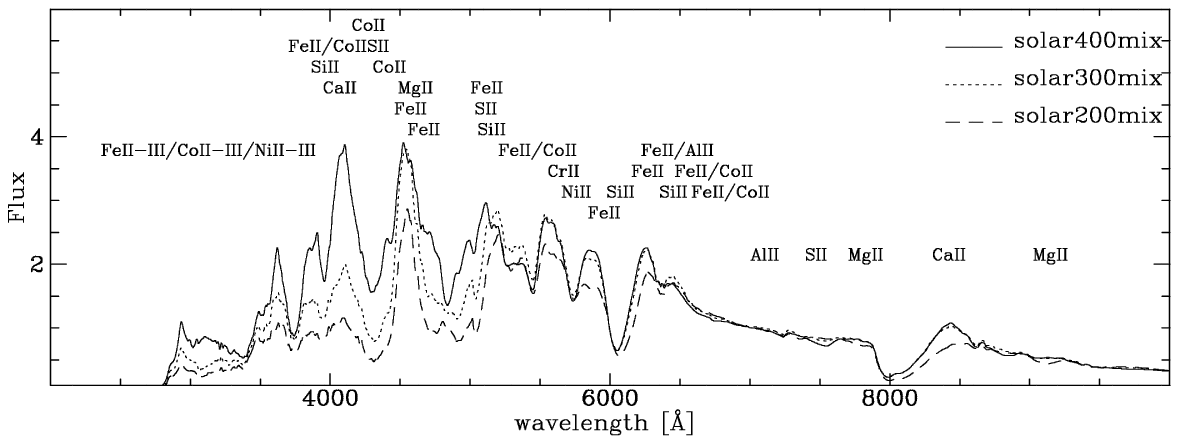,width=15.6cm,rwidth=9.5cm,clip=,angle=270}
\figure{9}{Comparison of synthetic spectra at maximum light assuming
complete mixing of the outer layers, involving
 64 \% (solar200mix),
 48 \% (solar300mix) and 24 \% (solar400mix) of the total mass, respectively 
(see Figs. 3 \& 4).
}
\endfig

\section{Implications for Cosmology}

 Systematic effects that must be taken into account in the use of SNe~Ia to
determine
cosmological parameters include technical problems, changes
of the environment with time,  changes in the statistical
 properties of the SNe~Ia, and changes in the physical properties of SNe~Ia
which are the main
subject of this study. The estimates  of the redshifts when evolutionary
effects become
important given below,
 are based on the assumptions (!)  of a flat universe with an age of 14 Gyrs,
and of evolution time scales for the progenitors of about 0.5 to 6 Gyrs which
are given by the
ZAMS life times  of Pop II
 donor stars with masses between 1 and 7 $M_\odot$ (Schaller et al. 1992).

 In the first class of technical problems,
 we put  corrections for redshift. If standard
filter systems are used, they can be well calibrated to local standards but the
k-correction is of some concern. This  problem can be overcome if redshifted
``standard" filters are used. Another
  technical problem  may arise then from the fact that the
transmission functions must be identical to those resulting from the redshift
because
no direct calibration can be applied by using a comparison star.

 In the second class of problems associated with enviroment,
an important example is that the properties of dust
may change at high redshift. In the first place,  the element abundances in the
ISM can
change. In addition,
important donors  of dust such as low mass red giant stars
cannot contribute  because their evolutionary time is comparable to or longer
than
the age of the universe at z $\approx 0.5$ to $1$. Another problem related to
the correction
for extinction is that
the extinction law to be applied depends on the redshift
of the absorbing dust cloud.  For details, see H\"oflich \& Khokhlov (1996).

 In the  third class of problems related to the statistical properties
 is the fact  that the contribution of different
progenitor types may change with redshift.
 For local supernovae, it is likely that a large variety of binary
star properties (total mass of donor stars, separation, etc.) with very
different
 evolutionary life times
 account for the variety of SNe~Ia observed (H\"oflich et al. 1997).
Given this variety, we  expect a time evolution of
the statistical properties of the progenitors because, early on, progenitors
with
a short life time will dominate the sample. Among other changes, more massive,
shorter lifetime progenitors
will  have a lower C/O ratio in the center  (see below). Changes in the
statistical properties
are expected to increase substantially at
     redshifts larger than 0.7 to 0.8 where the age of the universe becomes
comparable to or shorter than the suspected  progenitor life times.
 Other evolutionary effects have been
mentioned in the introduction: The ZAMS life time changes with metallicity,
Roche lobe radii change with
metallicity and  the lower limit for accretion and steady burning of hydrogen
 on the surface of the  WD
changes by a factor of 2 between Pop I and Pop II stars (Nomoto et al., 1982).

Finally, the properties of the typical, individual SNe~Ia may change. If more
massive stars contribute
to the supernova  population, we expect a smaller C/O ratio. Changing the
 C/O ratio from 1/1 to  2/3 with otherwise identical parameters (see \S
3.2),
 will result in a smaller \ni production and, consequently, a lower
bolometric luminosity at late times.
 On the other hand, the slower expansion causes less adiabatic cooling during
early times and, thus,
the luminosity at maximum light is larger (Fig. 6). This implies that the peak
 to tail luminosity
ratio  changes with the C/O ratio.
 The consequence  for analyses  that use light curve shapes or the postmaximum
decline
(e.g. $\delta m_{15}$) is evident. Quantitatively, for monochromatic light
curves in our example,
the difference in the peak to tail luminosity ratio
is $\approx 0.1~...~0.2 ^m$. For techniques to determine the absolute peak
luminosity
by measuring either  $\delta m_{15}$ or LC shapes (e.g. Hamuy et al. 1995,
Riess et al. 1995),
 this translates
into a systematic error of $\approx 0.3^m$ in $m_v$. This systematic change in
peak luminosity with
redshift due to changing C/O ratios could mimic the effects of cosmology.
Note that the transition density from
deflagration to detonation may depend on the energy release during the
deflagration phase and  hence
the transition density could be a function of the C/O ratio. Both effects could
alter the LC.

The other physical effect on the SNe~Ia themselves
 is the influence of the initial metallicity Z on the nuclear
 burning conditions during the explosion.  This
 produces a change of the isotopic composition in the outer layers. A reduction
of Z with redshift z
is expected. Note that the metallicity corresponds to the redshift when the
progenitor is formed and
not the redshift when the SN~Ia is observed.
 Although small, the changes in the spectrum with metallicity
 have an important effect on the colors of SN at high red shifts where they are
shifted into other
bands (see
Fig. 10). For local SNe~Ia, a change of the metallicity by a factor of 3
implies a variation in color of
two to three hundredths of
a magnitude. For $z \geq 0.2$,
however, Fig. 10 shows that
such a change can induce a systematic change in B-V of up to 0.3 magnitudes.
 For V-R and R-I, the effect remains small for redshifts z $\leq $ 0.5 and 0.7,
respectively, but it
increases to the same order at redshifts
relevant for the determination of cosmological parameters.
 The amplitude of this effect and hence the uncertainty in color, reddening and
brightness with metallicity
 is again comparable to the brightness change imposed
by cosmological deceleration.
 
\begfig 0.1cm
\psfig{figure=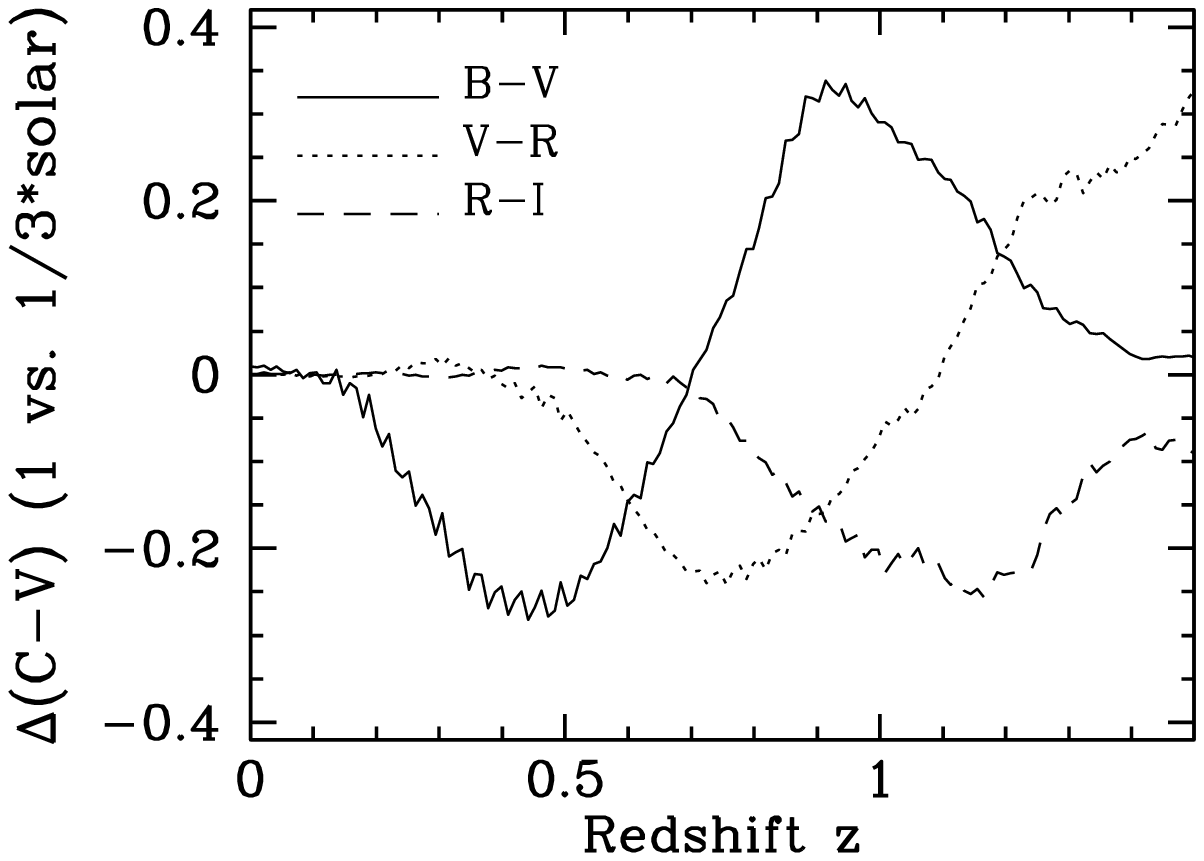,width=10.6cm,rwidth=6.5cm,clip=,angle=270}
\figure{10}{Metallicity effect on B-V and R-V for DD21c as a function of the
redshift.}
\endfig

 If SNe at moderate $\approx 0.5$ and  high red shifts ($\approx 1.0$)
are used to determine $\Omega _M$ and $\Omega_\Lambda $, respectively,
 these systematic shifts and changes of the
amplitude of the variation the influence of the metallicity on the color
indices
 may prevent a proper determination
of the cosmological parameters if it is not taken into account.
Due to the effect of metallicity on color
 and the different time scales for progenitor
 evolution and hence statistical properties  discussed above, the distribution
of peak brightness as a
function of light curve shape and redshift is likely to
produce  a scatter in the distribution
of the cosmological parameters.
 The way to reduce this scatter is to understand its physical origin in the
sorts of
effects we have outlined here.
Detailed spectral analyses of different SNe may show that the effects of
 metallicity  remains
moderate (e.g. because, even today, all progenitors may have a low Z),
or that it  can be corrected by using models. Quantitatively, other dynamical
 models and metallicities  will give different amplitudes
of $\Delta  (B-V)$,    $\Delta  (V-R)$, and $\Delta  (R-I)$
at given z than illustrated.
 In general, the use of red shifted filters will eliminate the major
systematic effect illustrated in Fig. 10 but, even
putting aside the systematic
problems due to the realization of identical transmission functions, the small
`noise'  (see Fig. 10)
due to the `bumpy' spectra is real and  requires a very narrow grid of filters
with respect to redshift, or the determinations
of $q_o$ will show a significant, intrinsic spread.  Even if corrections for
redshift
and the small wiggles are  applied and if corrections for the interstellar
redding are
determined,  a systematic effect of the order of $0.1^m$
will remain because of the small, but systematic dependencies
 in the color indices  at z=0 (see Fig. 10).

\begtab
\table{4}{Uncertainties in $q_o$, $\Omega _M$ and $\Omega_\Lambda$ as introduced by
composition effects}
\hline
\+~~~~~~~~~~~ $\Delta m  [mag]$ ~~~~~~~~~~~~~~~~~~~~~   &~~~ 0.1 ~~~& 
~~~ 0.2 ~~~ & ~~~ 0.3 ~~~ & ~~~ 0.6 ~~~ \cr
\hline
\+ $q_o(z=1)$                           & 0.1 & 0.2 & 0.3 &  0.6\cr
\+ $\Omega _M(z=1,\Omega_\Lambda =0)$          & 0.2 & 0.4 & 0.6 &  1.2\cr
\+ $\Omega _M(z=1.0,\Omega_\Lambda =0.5,flat)$ & 0.1 & 0.2 & 0.3 &  0.6\cr
\hline

\endtab

 In conclusion, both the change of Z and the
C/O ratio in the progenitor may each be expected to give variations in the peak
brightness
derived of the
order of a few tenth of a magnitude as one goes from z$\approx 0$ to z$\approx
$1.
  As a guideline, we give in Table 4 the size of the systematic effects
as they enter $q_o$ , $\Omega_M$ and $\Omega_\Lambda$  to first order for
 Friedmann-Lemaitre cosmological models (Goobar \& Perlmutter 1995,
 Perlmutter et al. 1997) for which
 $q_o=\Omega_M/2 -\Omega_\Lambda$ with $\Omega_\Lambda=\Lambda/(3~H_o^2)$ and,
for a flat univers, $\Omega_M +\Omega_\Lambda=1$.
To first order, the uncertainties are $\propto 1/z$ and $\Delta m$.
An uncertainty  $\Delta m=0.6^m$ should be regarded as worst case scenario.
 At this point, a warning is    appropriate.
 Although the assumed size of the change in the initial metallicity Z
 is of the right order  based on the chemical evolution of galaxies (Samland et
al. 1997)
 metallicity change as a function of galactic evolution is highly uncertain.
Yet another
uncertainty enters because Z is set by the time the progenitor is born and
there can be
a large range of times before the star dies.
 Moreover, the statistical  properties of progenitor systems are not well
known. For instance,
 the mass of the progenitors and the donor stars
must be in the range between $\approx 1 $ and $7 ~M_\odot $,
but the relative contributions within this mass range are unknown even for
local SNe~Ia.
 To beat down these uncertainties,  comparative analyses of the light curves
and spectra (e.g. velocities, lineshifts, spectral  flux distributions) of both
 the local and distant SNe~Ia are required.

\section{ Final Discussion and Conclusions}

 Using a delayed detonation model as an example, we have studied
the possible influence of the initial chemical composition on the light curves
and spectra
of  SNe~Ia.

 At early epochs
 the mean life time of progenitors will be smaller than at present
 and, hence, the mean progenitor mass
is larger compared to the current epoch (see \S  4).
 A substantially  reduced  C/O ratio in the inner region of the progenitor
white dwarf
will result if the WD originates from a star with
more than $\approx 3 - 4 M_\odot$ rather than 1  - 2 $M_\odot$
 on the ZAMS.
 We find that a reduction of the initial C/O ratio has effects similar to a
lower  transition
density from deflagration to detonation.
 In both cases, the location of the Si-rich material is shifted
to  smaller expansion velocity.  In the case of a reduction in C/O,
 the expansion velocity during the explosion is somewhat lower
and the time is delayed until the transition density is reached.
 For lower transition densities, the extent of the
 deflagration phase is increased because a longer time is
required to reach the transition density. In both cases,
this allows for a longer pre-expansion of the outer layers and,
consequently, the \ni production is reduced.
The major difference between a reduced C/O ratio and a lower transition density
is the reduced explosion energy (and expansion rate) in the former case.
 This causes the peak in the Si distribution to be more pronounced in velocity
space,
a slower rise to maximum light by about 3 days,  an increased peak luminosity
and, consequently,
a steeper decline after maximum light when the stored, thermal energy is
exhausted.
The use of light curve shapes or post-maximum decline rates alone may
lead to  systematic errors in the
estimates of the absolute brightness by several tenths of a magnitude.
 In principle, detailed coordinated
analyses of spectra and light curves provide a way to disentangle the effects
e.g.
of a change in the transition density and the C/O ratio and to
determine the peak brightness on a case by
case basis. Empirically, a correlation analysis of the LC shape and the
expansion rate provides
a tool to test for the range of C/O ratios realized in nature (see sect. 3.2).

 Changing the initial metallicity from Population I to II
reduces the progenitor life time and Roche lobe radii and hence changes
 progenitor evolution. It also
alters the isotopic composition of the outer layers. Especially important is
the
 increase in the $^{54}$Fe production with metallicity.
 The initial WD composition has been found to have rather small effects on the
overall LCs.
The \ni production and hence the bolometric and monochromatic  optical and IR
light curves
differ only by a few
hundredths of a magnitude. This change is almost entirely due to the small
change in the \ni
production and not due to a change in the opacities because the diffusion time
scales are governed
by the deeper layers where burning is complete.
 The time of maximum light in the Johnson filter system  changes by $\approx 1
$ day as Z is varied from
solar to 1/3 solar.
 The short wavelength  part of the spectrum at maximum light is
affected by a change in Z.
   This provides a direct test for the initial metallicity of local SNe if well
calibrated
spectra are available and, thus, may give a powerful tool to unravel the nature
(and lifetime) of
SNe~Ia progenitors. To do so, spectra around or prior to maximum light are
needed because at later times
the inner layers dominate the spectra where the spectrum is not sensitive to
the initial metallicity
(see sect. 3.3).
 Note that quantitatively, the influence of Z
is      model dependent because it
depends on the photon redistribution and optical depth at a given time,
i.e. on the density and abundance structure.
 Even the
 sign of the effect if the configuration is very different than the delayed
detonations.
 As an empirical test, a   differential comparison
 between maximum and post-maximum spectra of different SNe~Ia with similar
light curve shapes and
expansion velocities provides a valuable tool to probe for Z effects
(see  sect. 3.2).

In terms of distant SNe, we may expect less metals {\it and} lower C/O ratios
in the  past. For red-shifted
supernovae, the Z effect may produce  systematic changes of up to 0.2 to 0.3
magnitudes
 in B-V, V-R and R-I
for z $\ge $ 0.2, 0.5 and 0.7, respectively.
The amplitude of the maximum change and the phase with respect to z is somewhat
model dependent. In general,   the C/O ratio will decrease with redshift and
can be
expected to become significant at redshifts $\geq 0.7$ when the age of the
universe
is comparable to  the lifetime of the progenitors.
 The systematic effects due to changing
 Z and C/O may be comparable in amplitude to the effects
due to the deceleration of the universe.
Other evolutionary effects have been qualitatively discussed in \S      4 and
the introduction.
Because the properties of SNe~Ia may
change  with z, a  measurement of $q_o$, $\Omega _M $ and $\Lambda$ may be
highly biased.
 Determinations of $H_o$ will remain unaffected if it is based on a sufficient
large range
of light curves (e.g. B, V, R, I).

 Finally, we also want to mention the limitations of this study which
need to be overcome  in the future. First of all, only one specific set
of  models has been studied in detail. Although the  model parameters
have been chosen to allow for a representation of   ``typical" SNe~Ia,
more comprehensive studies and detailed fitting of actual observations
are needed.  The C/O ratio has been changed for the entire WD; however,
the   C/O ratio is   expected to change only  for the mass before binary
accretion begins  (e.g. $ 0.6 \leq M_{in} \leq 1.2 M_\odot$). A C/O ratio of
$\approx 1$ will result from the accreted material at the surface regardless of
the
mass of the primary of secondary star  (e.g.
Nomoto et al. 1984).  The final composition of the supernova ejecta which
undergo  partial burning depends mainly on the density during burning.
This, in turn, depends on the pre-expansion during the deflagration
phase because it lasts longer by an order of magnitude (i.e. about
1 - 2 seconds) compared  to the detonation phase. During the
deflagration phase, only the inner regions ($\approx 0.2 ~-~0.3
M_\odot$) are burned. Therefore,  it does not make much difference whether
the C/O ratio is lower on the inside and 1:1 on the outside due to accretion
 whether a globally reduced C/O ratio is considered.
Nevertheless, a study of the influence of realistic structures is
desired in order for the effects to be analyzed more quantitatively.

This study explored  some  effects that can be expected if we go
back in  time. Detailed comparisons with observations are ultimately
needed to address the effects of composition evolution or to probe for their
unimportance.
 To do so,  well-sampled spectral and light curve data must be available such
as currently
obtained by the Berkeley and CTIO groups. To isolate  the Z effects we would
suggest that
maximum spectra and postmaximum spectra be  taken routinely.
  To acquire a full understanding of SNe~Ia, detailed analyses of nearby (e.g.
Virgo or closer)
supernovae  are needed.
 Such a program is  justified, because
there is already
evidence in the current supernova sample for evolutionary effects
that  are not yet understood, but suggest that some
effects of varying composition effects are present.  We might naturally
expect a range in initial metallicities of SNe~Ia within spiral galaxies
that have ongoing star formation and between ellipticals and
spirals.  The  same is true for the C/O ratio which will
depend on the average age of the progenitor population.
  Branch et al. (1996) and Hamuy et al. (1996) have
shown that the mean peak brightness is dimmer in ellipticals
which lack a young population.
 Wang et al. (1997) have shown that
the peak brightness in the outer region of spirals is
similar to those found in ellipticals but, in the central region, both
intrinsically brighter and dimmer SNe~Ia occur. Both these effects may  involve
                a change in the C/O ratio. SNe~Ia with
a low C/O ratio have a short progenitor lifetime and are predicted to be
brighter.
   Another independent statistical test for the
influence of a changing C/O ratio may be to use the spread in the velocity,
light curve shape relation. If the C/O ratio is the dominant effect to
explain the spread in the properties of LCs, there should be a tight
relation between LC shape and expansion velocity in ellipticals and a
spread in spirals. Another approach to get insight into the local
SNe~Ia may be the use of $\delta $-Cephei distances to evaluate
the spread in the absolute brightness.

\bigskip
{\bf Acknowledgements:} We thank the referee, D. Branch, and K.  Nomoto
 for helpful discussions and useful
comments.
This research was supported in part by NSF Grant AST 9528110 and NASA Grant NAG
5-2888,
and a grant from the Texas Advanced Research Program, Grant  Ho1177/2-2 from
the German Science
Foundation (DFG), Grant 20-47252.96 of the Swiss Nationalfonds, and in its
final stages
 by NSF Grant PHY 94-07194 to the ITP at the University of California.
 The  calculations were done at the
High Performance Computer Facilities of the University of Texas at Austin.

\bigskip

\heading{References}

\journal {Abbot  D.C., Lucy  L.B.}{1985}{ApJ}{288}{679}

\journal {Arnett  W. D.}{1969}{Ap. Space Sci.}{5}{280}

\journal {Barbon  R., Ciatti  F., Rosino  L.}{1973}{A\&A}{29}{57}

\journal {Barbon  R., Benetti  S., Cappellaro  E., Rosino  L., Turatto  M.}
         {1990}{A\&A}{237}{79}

\inbook  {Baron  E., Hauschildt  P.H., Mezzacappa  A.}{1997}{Thermonuclear
Supernovae}
{P. Ruiz-Lapuente, R. Canal, J. Isern}{Nato ASI Series}{Kluwer Academic
Publisher, Vol. 486}{627}

\journal {Branch  D.}{1981}{ ApJ }{248}{1076}

\journal {Branch D., Tammann G.A}{1992} {ARA\&A}{30}{359}

\journal {Branch  D., Romanishin W., Baron E.}{1996}{ ApJ }{467}{73}

\inbook{ Canal  R.}{1994}{ Proc. of Les Houches Session
LIV}{ S. Bludman, R. Mochovitch, J. Zinn-Justin}{Paris}{North-Holland}{155}

\journal {Castor H.U. }{1974}{MNRAS} {169} {279}

\journal{Collela  P., Woodward  P.R.}{1984}{J.~Comp.~Phys.}{54}{174}

\inbook  {DiStefano R., Nelson L.A., Lee W., Wood T.H., Rappaport S.}{1997}
{Thermonuclear Supernovae}
 {P. Ruiz-Lapuente, R. Canal, J. Isern}{Nato ASI Series}{ Kluwer Academic
Publisher, Vol. 486}{147}

\journal {Filippenko  A.V., Richmond  M.W., Matheson  T., Shields  J.C.,
Burbidge  E.M., Cohen  R.D., Dickinson  M., Malkan  M.A., Nelson  B.,
Pietz  J., Schlegel  D., Schmeer  P., Spinrad  H., Steidel  C.C.,
Tran  H.D., Wren  W.} {1992a} {ApJ} {384} {L15}

\journal{Filippenko  A.V., Richmond  M.W., Branch  D., Gaskell  C.M.,
         Herbst  W., Ford  C.H., Treffers  R.R., Matheson  T., Ho  L.C.,
          Dey  A., Sargent  W.L., Small  T.A., van Breugel  W.J.M.}
         {1992b} {AJ}{104}{1543}

\journal  {Goodbar A., Perlmutter S.}{1996}{ApJ}{450}{14}

\journal  {Hamuy M., Phillips M.M, Maza J., Suntzeff N.B., Schommer R.A.,
Aviles A.} {1996}
 {AJ} {112}{2438}

\journal {Hansen  C. J., Wheeler J. C.}{1969}{Ap. Space Sci.}{3}{464}

\inbook  {Harkness R.P.}{1991} {ESO/EIPC Workshop. SN1987A and Other
Supernovae}
 {I.J. Danziger \& K. Kj="ar}{ESO}{Garching}{447}

\inbook{Hernanz M., Salaris M., Isern J., Jose J.}{1997}
 {P. Ruiz-Lapuente, R. Canal, J. Isern}{Nato ASI Series}{Kluwer Academic
Publisher}{Vol. 486}{167}

\book{H\"oflich, P.}{1990}{Habilitation Thesis, Ludwig Maximilians Univ.}
{M\"unchen}{published as MPA-90 563}

\journal {H\"oflich  P., Khokhlov  A., M\"uller  E.} {1992} {A\&A} {259} {243}

\journal {H\"oflich  P., Khokhlov A., M\"uller E.} {1992} {A\&A} {259} {243}

\journal {H\"oflich  P., M\"uller  E. \&  Khokhlov  A.} {1993} {A\&A} {268}
{570}

\journal {H\"oflich P.}{1995}{ApJ}{443}{533}

\inbook {H\"oflich P., M\"uller E., Khokhlov A., Wheeler J.C.}{1995b}{$17^{th}$
Texas Symposium on
Relativistic Astrophysics}{Tr\"umper}{Munich}{Springer}{348}

\inbook{H\"oflich P., Khokhlov A., Nomoto K., Thielmann F.K., Wheeler
J.C.}{1997}
 {P. Ruiz-Lapuente, R. Canal, J. Isern}{
Nato ASI Series}{ Kluwer Academic Publisher}{Vol. 486}{659}

\journal {H\"oflich, P., Khokhlov A.} {1996} {ApJ} {457}{500}

\journal {Hoyle  F.,   Fowler  W. A.}{1960}{Ap. J.}{132}{565}

\journal{ Iben  I.Jr., Tutukov  A.V.}{1984}{ApJS}{54}{335}

\journal{ Iben  I.Jr.}{1997}
 {P. Ruiz-Lapuente, R. Canal, J. Isern}{Nato ASI Series,  Kluwer Academic
Publisher, Vol. 486}{111}

\journal {Ivanova  I. N., Imshennik  V. S.,     Chechetkin  V. M.}{1974}
          {ApSS}{31}{497}

\journal {Karp  A.H., Lasher  G., Chan  K.L., Salpeter E.E.}
{1977}{ApJ}{214}{161}

\journal {Khokhlov  A.}{1991}{A\&A} {245} {114}

\journal {Khokhlov  A., M\"uller  E.,   H\"oflich, P.} {1993} {A\&A} {270}
{223}

\book{Kurucz R.P.}{1993a}{CD-ROM 1:Line list}{CfA}{Harvard University}

\book{Kurucz R.P.}{1993b}{CD-ROM 23:Line list}{CfA}{Harvard University}

\book{Kurucz R.P.}{1996}{CD-ROM:Lines and Levels for iron group
elements}{CfA}{Harvard University}

\journal {Leibundgut  B., Kirshner  R.P., Filippenko  A.V.,
          Shields  J.C., Foltz  C.B., Phillips  M.M., Sonneborn  G.}
          {1991}{ApJ}{371}{L23}

\journal {Leibundgut  B. et al.}{1995}{ESO-Messenger}{81}{19L}

\journal {Livne E., Arnett D.} {1995}   {ApJ } {452}{62}

\book {Mihalas D.}{1978}{Stellar atmospheres}{Freeman}{San Francisco}

\journal {Mihalas D., Kunacz R.B., Hummer D.G.} {1975}   {ApJ } {202}{465}

\journal {Mihalas D., Kunacz R.B., Hummer D.G.}{1976}   {ApJ } {206}{515}

\journal {M\"uller  E.,   H\"oflich  P.}{1994} {A\&A} {281} {51}

\journal {Nomoto  K., Sugimoto  D.}{1977}{PASJ}{29}{ 765}

\inbook {Nomoto K.} {1980} {IAU-Sym. 93}{D. Sugimoto, D.Q. Lamb \& D.
Schramm}{Dordrecht}{Reidel}{295}

\journal { Nomoto  K.}{1982}{ApJ}{253}{ 798}

\journal{Nomoto  K., Thielemann  F.-K., Yokoi  K.}{1984}{ApJ}{286}{ 644}

\inbook { Nomoto  K., Yamaoka  H., Shigeyama  T., Iwamoto  K.}{1995}{Supernovae
and Supernova Remnants}{R. A. McCray, Z. Wang \& Z. Li}{Cambridge}{Cambridge
University
Press}{ 49}

\inbook { Nomoto  K., Yamaoka  H., Shigeyama  T., Iwamoto  K.}{1997}
{Thermonuclear Supernovae}
{P. Ruiz-Lapuente, R. Canal, J. Isern}{Nato ASI Series}{Kluwer Academic
Publisher, Vol. 486}{349}

\journal {Norgaard-Nielsen  H.U., Hansen L., Henning E.J., Salamanca A.A.,
 Ellis R.S., Warrick, J.C.}{1989}{Nature}{339}{523}

\journal {Nugent P., Baron E.,Hauschildt P., Branch D.}{1995}
{ApJ Let.}{ 441}{L33}

\infuture {Nugent P., Baron E.,Hauschildt P., Branch D.}{1997}
{ApJ}{in press}

\journal {Olson G.L., Auer L.H., Buchler J.R.} {1986}{JQSRT}{35}{ 431}

\inbook {
Paczy\'nski  B.}{1985}{Cataclysmic Variables and Low-Mass X-Ray
Binaries}{D.Q. Lamb, J. Patterson}{ Reidel}{Dordrecht}{ 1}

\circular {Pennypacker  C. et al.}{1991}{IAUC}{5207}

 \journal  {Perlmutter C. et al.}{1995}{ApJ Let} {440}{91}

 \journal  {Perlmutter C. et al.}{1997}{ApJ} {483}{565}

\journal {Phillips  M. M.}{1993}{ApJ}{413}{L108}

\journal{Phillips  M.M., Phillips  A.C., Heathcote  S.R., Blanco  V.M.,
         Geisler  D., Hamilton  D., Suntzeff  N.B., Jablonski  F.J., Steiner
         J.E., Cowley  A.P., Schmidtke  P., Wyckopf  S., Hutchings  J.B.,
         Tonry  J., Strauss  M.A., Thorstensen  J.R., Honey  W., Maza  J.,
         Ruiz  M.T., Landolt  A.U., Uomoto  A., Rich  R.M., Grindlay  J.E.,
         Cohn  H., Smith  H.A., Lutz  J.H., Lavery  R.J., Saha  A.}{1987}
        { PASP}{90}{592}

\journal {Pskovskii  Yu.P.}{1970}{Astron. Zh.}{47}{994}

\journal {Pskovskii  Yu.P.} {1977} {Sov. Astr.} {21} {675}

\journal {Rappaport S., Chiang E., Kallman T., Malina R.
}{1994a}{ApJ}{431}{237}

\journal {Rappaport S., DiStefano R., Smith J. }{1994a}{ApJ}{426}{692}

\journal {Riess A.G., Press W.H., Kirshner R.P.}{1995}{ApJ}{438}{L17}

\journal {Samland M., Hensler G., Theis C.}{1997}{ApJ}{476}{544}

\journal {Sandage  A., Tammann  G.A.}{1993} {ApJ} {415}{1}

\journal {Schaller G., Schaerer D., Meynet G., Maeder A.}{1992}{A\&A
Suppl}{96}{269}

\journal {Schmidt B.P, et al.}{1996}{BAAS}{189}{1089}

\journal {Sobolev, V.V.}{1957}{Sov. Astron.}{1}{297}

\inbook{Thielemann F.K., Nomoto K., Iwamoto K., Brachwit, F. }{1997}
 {P. Ruiz-Lapuente, R. Canal, J. Isern}{Nato ASI Series}{ Kluwer Academic
Publisher}{Vol. 486}{485}

\journal{Thielemann  F.-K., Nomoto  K., Hashimoto  M.}{1996}{ApJ}{460}{408}

\inbook{ Thielemann  F.-K., Arnould  M., Truran  J.W.}{1987}
{ Advances in Nuclear Astrophysics} {E. Vangioni-Flam}{Editions
fronti\`eres}{Gif sur Yvette}{525}

\journal{Thielemann  F.-K., Nomoto  K., Yokoi  K.}{1986}{A\&A}{158}{17}

\journal {Tsvetkov, D.Yu} {1994}{Astron.L.}{20}{374}

\journal {Van den Heuvel E.P.J., Bhattacharya D., Nomoto K., Rappaport S.}
{1992}{A\&A}{262}{97}

\journal {Wang L., H\"oflich P., Wheeler J.C. }{1997}{ApJ}{483}{29}

\journal{Webbink  R.F.}{1984}{ApJ}{277}{355}

\journal {Wheeler  J. C., Harkness  R .P.}{1990}{Rep. Prog. Phys.}{53}{1467}

\journal  {Wheeler J. C., Harkness R .P., Khokhlov A., H\"oflich P.}
{1995}{Phys. Rep.}{256}{  211}

\journal {Woosley  S.E., Weaver  T.A.}{1986}{ARAA}{24}{}

\infutbook{ Woosley  S. E. \& Weaver , T. A.}{1994}
{Proc. of Les Houches Session
LIV}{ S. Bludman, R. Mochovitch, J. Zinn-Justin}{Paris}{North-Holland}{63}

\journal {Woosley  S. E. \& Weaver, T. A.} {1994}{Ap. J.} {423}{371}

\inbook {Woosley S. E., Weaver T.A., Taam R.E. } {1980} {in: Type I
Supernovae}{J.C.Wheeler}{Austin}{U.Texas}{96}

\journal {Yamaoka H., Nomoto K., Shigeyama T., Thielemann
F.}{1992}{ApJ}{393}{55}

\bigskip
\end

\vfill\eject

\baselineskip=24pt
\begtab
\table{1}
{ Basic parameters for 
the  delayed detonation models.
}
 
\hline 
\+Model~~~~~ &  $M_\star$ ~~~~~& $\rho _c$ ~~~~~~~~~ & $\alpha$ ~~~~~&   $\rho_{tr}$~~~~~~~~~~~~ & ~~C/O~~ & ~~$R_Z$~~ & $E_{kin}$~~~~~  & $M_{Ni}$~~~~~  \cr
%
\hline                     
\+DD13c   & 1.4  &  2.6   &  0.03  &  3.0 & 1/1 & 1/1 & 1.36 &
                           0.79  \cr
\+DD21c   &  1.4  &  2.6   &  0.03  &  2.7 & 1/1 & 1/1  &  1.32  &
                           0.69  \cr
\+DD23c   &  1.4  &  2.6   &  0.03  &  2.7 & 2/3 & 1/1 &  1.18  &
                           0.59  \cr
\+DD24c   &  1.4  &  2.6   &  0.03  &  2.7 & 1/1 & 1/3  &  1.32  &
                           0.70  \cr
\+DD25c   &  1.4  &  2.6   &  0.03  &  2.7 & 1/1 & 3/1  &  1.32  &
                           0.69  \cr
\+DD26c   &  1.4  &  2.6   &  0.03  &  2.7 & 1/1 & 1/10  &  1.32  &
                           0.73  \cr
\+DD27c   &  1.4  &  2.6   &  0.03  &  2.7 & 1/1 & 10/1  &  1.32  &
                           0.69  \cr
\hline         
\endtab

\vfill\eject

\baselineskip=24pt

\vfill\eject

\vfill\eject
\baselineskip=24pt

\vfill\eject
{\bf Figure Captions}

 \begfig 0.1cm
\figure{1}{ Influence of the metallicity on the structure of a WD. The shaded
band   marks
the region which allows for a successful reproduction of observed LCs and
spectra for SNe~Ia
(H\"oflich \& Khokhlov 1996). In the upper plots,
 the M-R relation  and the $\rho_c$-M relations are given
for solar metallicities and, in the lower plots, the difference (in \%) for
models with 1/100 of solar metallicity are given.
 On the scales of the upper plots, the curves corresponding to different
metallicities would merge.
\endfig
\noindent
\begfig 0.1cm
\figure{2}{Test for the consistency of the integrated luminosity based on
the solution of the monochromatic  and frequency integrated radiation transport
equation
for model DD21c (Table 1). About       780 frequencies have been used in the
former case.
 Note that this inconsistency only enters the Eddington factors
which are based on the solution of monochromatic transport equation, but
not the energy conservation.}
\endfig

\begfig 0.1cm
\figure{3}{ Density and velocity
as a function of  mass for three delayed detonation models (Table 1).\hfill}
\endfig
\begfig 0.1cm
\figure{4}{ Abundances as a function of the final expansion velocity for the
 three delayed detonation models of Figure 3.
Both the initial $^{56}Ni$ and the final Fe profiles are shown.}
\endfig

\begfig 0.1cm
\figure{5}{Abundances of different isotopes as a function of the
expansion velocity for models
DD21c, DD24c and DD27c that have initial compositions of
solar, 1/3 solar, and 1/10 of solar, respectively.}
\endfig

 \begfig 0.1cm
\figure{6}{Comparison of bolometric light curves of the delayed detonation
models DD21c, DD23c and DD24c with otherwise identical parameters but with
different
C/O ratios and metallicity relative to solar (C/O; $R_Z$) of (1;1), (2/3;1) and
(1;0.3),
respectively.
}
\endfig

\begfig 0.1cm
\figure{7}{Comparison of  light curves in B (left) and V (right) of the delayed
detonation
models DD21c, DD23c and DD24c with otherwise identical parameters but different
different
C/O ratios and metallicities relative to solar (C/O; $R_Z$) of (1;1), (2/3;1)
and (1;0.3).
 The difference relative to DD21c and the monochromatic LCs are given
in the upper and lower plots, respectively.
}
\endfig
\begfig 0.1cm
\figure{8}{Comparison of synthetic NLTE spectra at maximum light for
 initial compositions of solar and 1/3 of solar, respectively
(upper graph). The standard Johnson filter functions for UBV, and R are
also shown. In the lower graph, the radius as a function of
wavelength  where the monochromatic optical depth reaches
0.1 and 1, respectively  is given for DD21c.}
\endfig

\begfig 0.1cm
\figure{9}{Comparison of synthetic spectra at maximum light assuming
complete mixing of the outer layers, involving
 64 \% (solar200mix),
 48 \% (solar300mix) and 24 \% (solar400mix) of the total mass, respectively
(see Figs. 3 \& 4).
}
\endfig

\begfig 0.1cm
\figure{10}{Metallicity effect on B-V and R-V for DD21c as a function of the
redshift.}
\endfig
\vfill\eject
\vskip 8.cm
 \begfig 0.1cm
\psfig{figure=fig0.eps,width=15.6cm,rwidth=9.5cm,clip=,angle=270}
\figure{1}{ }
\endfig
\vfill\eject
\vskip 8.cm
\noindent
\begfig 5.1cm
\psfig{figure=fig01.eps,width=11.6cm,rwidth=6.5cm,clip=,angle=270}
\figure{2}{}
\endfig
\vfill\eject

\vskip 8.cm
\begfig 8.1cm
\psfig{figure=fig1.eps,width=15.6cm,rwidth=9.5cm,clip=,angle=270}
\vskip 1.cm
\figure{3}{}
\endfig
\vfill\eject
\begfig 5.1cm
\psfig{figure=fig2.eps,width=15.6cm,rwidth=9.5cm,clip=,angle=270}
\vskip  10.5cm
\figure{4}{ }
\endfig
\vfill\eject
\vskip 8.cm

\begfig 5.1cm
\psfig{figure=fig3.eps,width=15.6cm,rwidth=9.5cm,clip=,angle=270}
\figure{5}{}
\endfig
\vfill\eject
\vskip 8.cm

 \begfig 5.1cm
 \psfig{figure=fig7.eps,width=15.6cm,rwidth=6.5cm,clip=,angle=270}
\figure{6}{
}
\endfig
\vfill\eject

\begfig 5.1cm
\psfig{figure=fig8.eps,width=15.6cm,rwidth=9.5cm,clip=,angle=270}
\figure{7}{}
\endfig
\vfill\eject
\begfig 5.1cm
\psfig{figure=fig4.eps,width=15.6cm,rwidth=9.5cm,clip=,angle=270}
\figure{8}{}
\endfig
\vfill\eject

\begfig 8.1cm
\psfig{figure=fig6b.eps,width=15.6cm,rwidth=9.5cm,clip=,angle=270}
\figure{9}{}
\endfig
\vfill\eject

\begfig 8.1cm
\psfig{figure=fig5.eps,width=15.6cm,rwidth=9.5cm,clip=,angle=270}
\figure{10}{}
\endfig
\end